\documentclass{jfm}
\usepackage{graphicx}
\usepackage{epstopdf, epsfig}
\usepackage[usenames]{color}
\usepackage{amsmath}
\usepackage{array}
\newcommand\Wen{\mbox{\textit{We}}}
\newcommand\Ohn{\mbox{\textit{Oh}}}
\newcommand\Stn{\mbox{\textit{St}}}

\begin{document}
\title{Drop impact on superheated surfaces: short-time dynamics and transition to contact}
\author{Pierre Chantelot\aff{1,\corresp{\email{p.r.a.chantelot@utwente.nl}}} and Detlef Lohse\aff{1,2}}
\affiliation{\aff{1}Physics of Fluids Group, Max Planck Center for Complex Fluid Dynamics, MESA+ Institute, and J. M. Burgers Center for Fluid Dynamics, University of Twente, P.O. Box 217, 7500AE Enschede, Netherlands
\aff{2}Max Planck Institute for Dynamics and Self-Organisation, Am Fassberg 17, 37077 Gottingen, Germany}
\maketitle

\begin{abstract}
  When a volatile drop impacts on a superheated solid, air drainage and vapor generation conspire to create an intermediate gas layer that delays or even prevents contact between the liquid and the solid.
  In this article, we use high-speed synchronized reflection interference and total internal reflection imaging to measure the short-time dynamics of the intermediate gas film and to probe the transition between levitation and contact.
  We observe that the substrate temperature strongly affects the vertical position of the liquid-gas interface and that the dynamic Leidenfrost transition is influenced by both air and vapor drainage (\emph{i.e}, gas drainage), and evaporation, the later giving rise to hitherto unreported vertical oscillations of the gas film that can trigger liquid-solid contact.
  We first derive scaling relations for the height of the gas film trapped under the drop's centerline, called the dimple height, and the minimum film thickness at short times. The former is set by a competition between gas drainage and liquid inertia, similarly as for isothermal impacts, while the later strongly depends on the vapor production.
  The gas pressure, at the location where the minimum thickness is reached, is determined by liquid inertia and vapor production and ultimately balanced by the increasing interfacial curvature, determining the minimum thickness.
  We show that, in the low impact velocity limit, the transient stability of the draining gas film remarkably makes dynamic levitation less demanding than static levitation. We characterise the vertical gas film oscillations by measuring their frequency and monitoring their occurrence in the parameter space spanned by surface temperature and impact velocity. Finally, we model the occurrence of these oscillations and account for their frequency through an hydrodynamic mechanism.
\end{abstract}


\section{Introduction}
The impact of a liquid drop on a solid target has been extensively studied \citep{yarin2006,josserand2016,tropea2017} since Worthington's pioneering observations \citep{worthington1877}.
The dynamics of the spreading liquid can now be described using analytical expressions once the drop touches the substrate \citep{gordillo2019}.
The initial stage of spreading is governed by inertia so that the wetted radius follows Wagner's theorem \citep{riboux2014}. Later, viscous and capillary forces oppose the outwards motion and ultimately balance the initial inertia as the maximal radius is reached \citep{laan2014,wildeman2016}.
Yet, before spreading starts, the air trapped between the falling drop and the substrate must drain for contact to occur.
This phase, during which the drop levitates on an air cushion, can affect the outcome of impact as revealed by the dramatic influence of ambient pressure on splashing \citep{xu2005,driscoll2011}.
This levitation phase can even dominate and last during the whole impact for low impact velocities \citep{kolinski2014_2}.
This holds even more when the substrate is heated far above the boiling point of the liquid, as then the impacting drop is separated from the substrate by a cushion of its own vapor, a situation called the dynamic Leidenfrost effect \citep{tran2012}.

The Leidenfrost effect \citep{leidenfrost1756,quere2013} has received a lot of attention owing to its relevance in heat transfer applications such as spray cooling \citep{kim2007,breitenbach2018}, the cooling of fuel rods in case of a nuclear incident \citep{hamdan2015}, or spray combustion \citep{moreira2010}.
As the presence of gas between the liquid and the solid dramatically reduces the heat flux, phase diagrams distinguishing levitation and contact behaviors have been obtained for many liquid-solid combinations based on side-view imaging \citep{yao1988,bernardin2004} and, more recently, on the direct bottom-view observation of contact \citep{tran2012,staat2015,shirota2016}.
Understanding the dynamics of the air and vapor film (together referred to as gas from now on) trapped at impact on superheated substrates is thus key to predict the heat transfer.

When the liquid and the solid are both at ambient temperature, the drainage of the air film trapped between the falling drop and the substrate has been thoroughly investigated.
In this situation, that we call an isothermal impact, the pressure buildup under the drop results in the formation of a dimple whose edge is marked by a localized region of high curvature, called the neck \citep{mandre2009,hicks2010,mani2010,bouwhuis2012}.
This region eventually moves downwards, leading either to the wetting of the substrate and the entrapment of an air bubble \citep{chandra1991,thoroddsen2005}, or, for low enough impact velocities and sufficiently smooth substrates, to the creation of a relatively flat air film that can enable drop bouncing \citep{kolinski2014_2}.
In the later case, the dimple and neck have a fixed radial position and height as the liquid spreads \citep{kolinski2012,kolinski2014}.
For impacts on superheated substrates, however, it is yet unclear how evaporation contributes to the evolution of the gas film, although it is already known that the dimple and neck structure also appears \citep{shirota2016}.

In this article, we aim to disentangle the effects of gas drainage and vapor generation on the short-time drop impact dynamics to provide insight into the transition from drop levitation to contact. In order to do so, we probed the dynamics of the gas layer squeezed between the liquid and the substrate at impact and monitored the failure of the gas film. We observed hitherto unreported vertical oscillations of the drop bottom interface, specific to superheated conditions, that can trigger contact events. We thus seek to understand the role of these oscillations on the transition towards the dynamic Leidenfrost effect.
The experimental setup and procedure is detailed in \textsection 2. In \textsection 3, we discuss the evolution of the film shape at short-time and the influence of surface temperature on its characteristic features. We also report the measurements of the minimum thickness at short times and provide a model to account for them.
In \textsection 4, the transition from levitation to contact is discussed, based on identifying the type of collapse of the gas layer in the parameter space spanned by the surface temperature and the impact velocity.
Finally, we model the vertical oscillations of the drop's base and discuss their role in the transition towards the dynamic Leidenfrost effect.
The paper ends with conclusions and an outlook (\textsection 5).

\section{Experimental setup and procedure}
\begin{figure}
  \centering
  \includegraphics[width=1\textwidth]{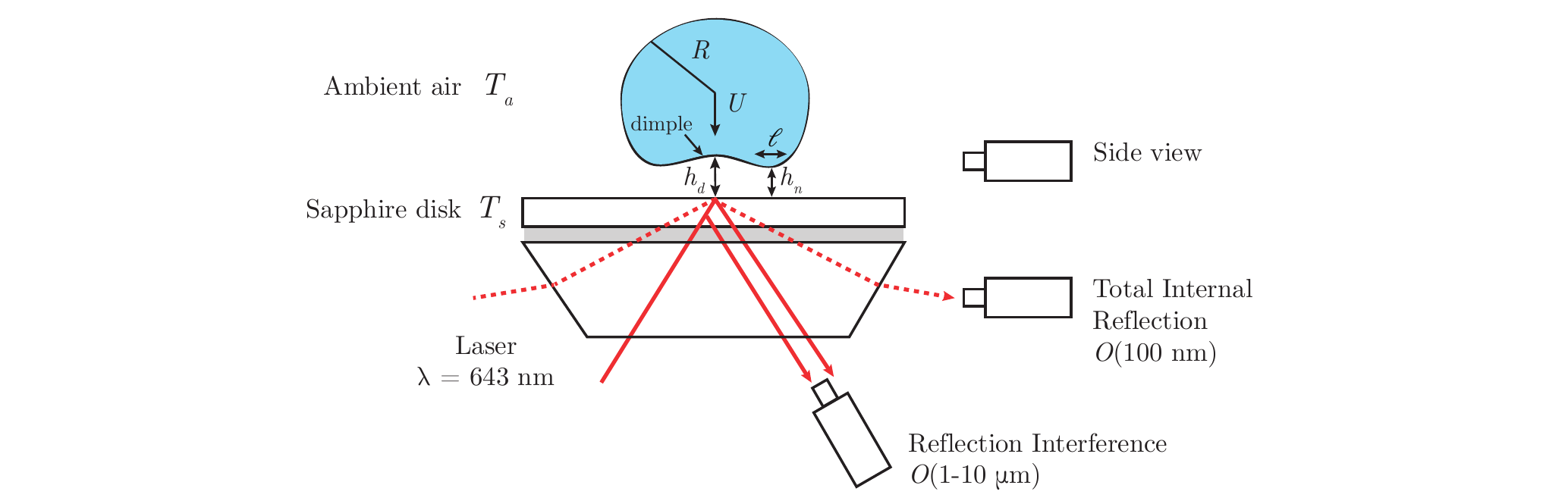}
  \caption{Ethanol drops with equilibrium radius $R$ and velocity $U$ impact a heated sapphire substrate with temperature $T_s$. We record side views and use reflection interference (RI) and total internal reflection (TIR) imaging to measure the thickness of the gas film squeezed between the liquid and the solid with three synchronized high-speed cameras. We define in the sketch the dimple and neck height, $h_d$ and $h_n$, respectively, as well as the radial extent of the neck region $\ell$ (sketch not to scale).}
\label{fig0}
\end{figure}

Our experiments, sketched in figure \ref{fig0}, consist in impacting ethanol drops on an optically smooth sapphire disk.
We choose this liquid-solid combination as (i) it allows us to neglect vapor cooling effects during impact and (ii) the excellent thermal conductivity of sapphire ($k_s = 35$ W/K/m) approximates isothermal substrate conditions \citep{vanlimbeek2016, vanlimbeek2017}.
The sapphire substrate is coupled to a glass dove prism using silicone oil and mounted in an aluminium heating block. The substrate temperature $T_s$ is set to a fixed value between $105$$^\circ$C and $270$$^\circ$C using a propoportional-integral-derivative controller, and measured with an external surface probe.
From this, we deduce the superheat $\Delta T = T_s - T_b$, where $T_b = 78$$^\circ$C is the boiling temperature of ethanol.
Drops with radius $R = 1.0 \pm 0.1$ mm are released from a calibrated needle, whose height is adjusted to obtain impact velocities $U$ ranging from 0.3 m/s to 1.2 m/s.
The chosen drop radii and impact velocities correspond to a Reynolds number $\Rey = \rho_l R U/\eta_l$ and an Ohnesorge number $\Ohn = \eta_l/\sqrt{\rho_l R \gamma}$, with typical values of 500 and 0.01, respectively, \emph{i.e} initially viscous effects can be neglected compared to inertia and capillarity. The Weber number $\Wen = \rho_l RU^2/\gamma$, ranging from 3 to 50, further indicates that experiments are conducted in the inertial regime. Table \ref{table1} gives an overview of the properties of the liquid and of the two components of the gas phase: air (with ambient temperature $T_a$) and ethanol vapor.
Note that the material properties of the fluids are temperature dependant and the temperature at which they should be evaluated will be discussed throughout the manuscript.

\begin{table}
\centering
\begin{tabular}{c p{0.35\textwidth} >{\centering\arraybackslash}p{0.18\textwidth} >{\centering\arraybackslash}p{0.18\textwidth} >{\centering\arraybackslash}p{0.18\textwidth}}
 & \textit{Description} & Ethanol (\textit{l}) & Ethanol (\textit{v}) & Air \vspace{4pt}\\
$-$ & Temperature ($^\circ$C) & 20 & 80 & 20 \\
$\rho$ & density (kg/m$^3$) & 789 & 1.63 & 1.2 \\
$\eta$ & viscosity (mPa.s) & 1.2 &  $1.05 \times10^{-2}$ & $1.85 \times10^{-2}$ \\
$C_p$ & specific heat (kJ/kg/K) & 2.4 & 1.8 & 1.0 \\
$k$ &  thermal conductivity (W/K/m) & 0.171 & 0.023 & 0.026 \\
$\kappa$ & thermal diffusivity (m$^2$/s) &  $0.09 \times 10^{-6}$ & $7.8 \times 10^{-6}$ & $21.7 \times 10^{-6}$ \\
$\mathcal{L}$ & latent heat (kJ/kg) & 853 & \rule{0.4cm}{0.5pt} & \rule{0.4cm}{0.5pt} \\
$\gamma$ & surface tension (N/m) & 0.022 & \rule{0.4cm}{0.5pt} & \rule{0.4cm}{0.5pt} \\
\end{tabular}
\caption{\label{table1} Physical properties of ethanol in the liquid (\textit{l}) and vapor (\textit{v}) phase and of air. }
\end{table}

We study the impact dynamics using three synchronized high-speed cameras to obtain side views and interferometric measurements of the gas film (figure \ref{fig0}). We record side views at 20 000 frames per second (Photron Fastcam SA1.1) from which we determine the drop radius $R$ and the impact velocity $U$.
We measure the gas film thickness profile by coupling two interferometry techniques that have been successfully applied in the context of drop impact: single-wavelength reflection interference \citep{driscoll2011,li2015,staat2015} and total internal reflection imaging \citep{kolinski2012,shirota2016}.
Simultaneously using these two techniques allows us to realise the benefits of both of them: reflection interference (RI) enables relative thickness measurements of thin films up to tens of micrometers while total internal reflection imaging (TIR) provides an absolute information on the evanescent lengthscale on the order of 100 nm. The combination of both techniques gives access to the absolute thickness of the entire profile of the gas layer during drop impact.

We implement these two techniques simultaneously by expanding a diode laser spot with wavelength $\lambda = 643$ nm into a collimated beam that we split into two optical paths leading to the substrate.
The beam used for RI imaging is introduced through the bottom face of the dove prism at a slight angle to only observe interferences generated by reflections at the top of the substrate and the bottom of the drop. The interference patterns are composed of dark and bright fringes that we image using a long distance microscope (Navitar telecentric 12x) mounted on a high-speed camera (Photron SA-Z) to obtain a typical resolution of 8 $\mu$m/px at a frame rate of 700 kHz.
The TIR beam is p-polarized and introduced through one of the sloped faces of the prism so that it  undergoes total internal reflection at the top of the substrate. The interaction of the emitted evanescent wave and the impinging drop results in a decrease of the reflected intensity that enables the measurement of the film thickness. When the drop touches the surface, light transmits through the liquid creating a sharp change in grayscale intensity that allows to distinguish wetted from dry areas.
The resulting images are recorded at 225000 frames per second (Photron Nova S12) with a long distance microscope with resolution $10$ $\mu$m/px.
The details of image processing and calibration of interferometric measurements are further discussed in appendix \ref{appendix_setup}.

\section{Short-time dynamics of the gas film}
Side views of the impact event of an ethanol drop in the dynamic Leidenfrost regime are presented in figure \ref{fig1}a ($U = 0.5$ m/s, $\Wen = 9.9$ and $T_s = 164$$^\circ$C). The impinging liquid spreads, recoils and finally detaches from the substrate after a time on the order of 10 ms.
This time, that we call the rebound time, is set by the inertio-capillary timescale $\tau_c = \sqrt{\rho_lR^3/\gamma}$ \citep{richard2002,biance2006}.
Here we define the origin of time ($t=0$) as the time of the first frame where the drop enters within the evanescent lengthscale.
We focus on the dynamics of the gas film squeezed between the liquid and the hot substrate at the first instant of impact, that is for $0 < t < \tau_i$ where $\tau_i = R/U$ is the inertial timescale.
For a drop with radius $1$ mm impacting at $1$ m/s, this inertial time is $\tau_i = 1$ ms and therefore much shorter than the inertio-capillary time $\tau_c \approx 10$ ms.
\subsection{Phenomenology}
\subsubsection{Sequence of events}
\begin{figure}
  \centering
  \includegraphics[width=1\textwidth]{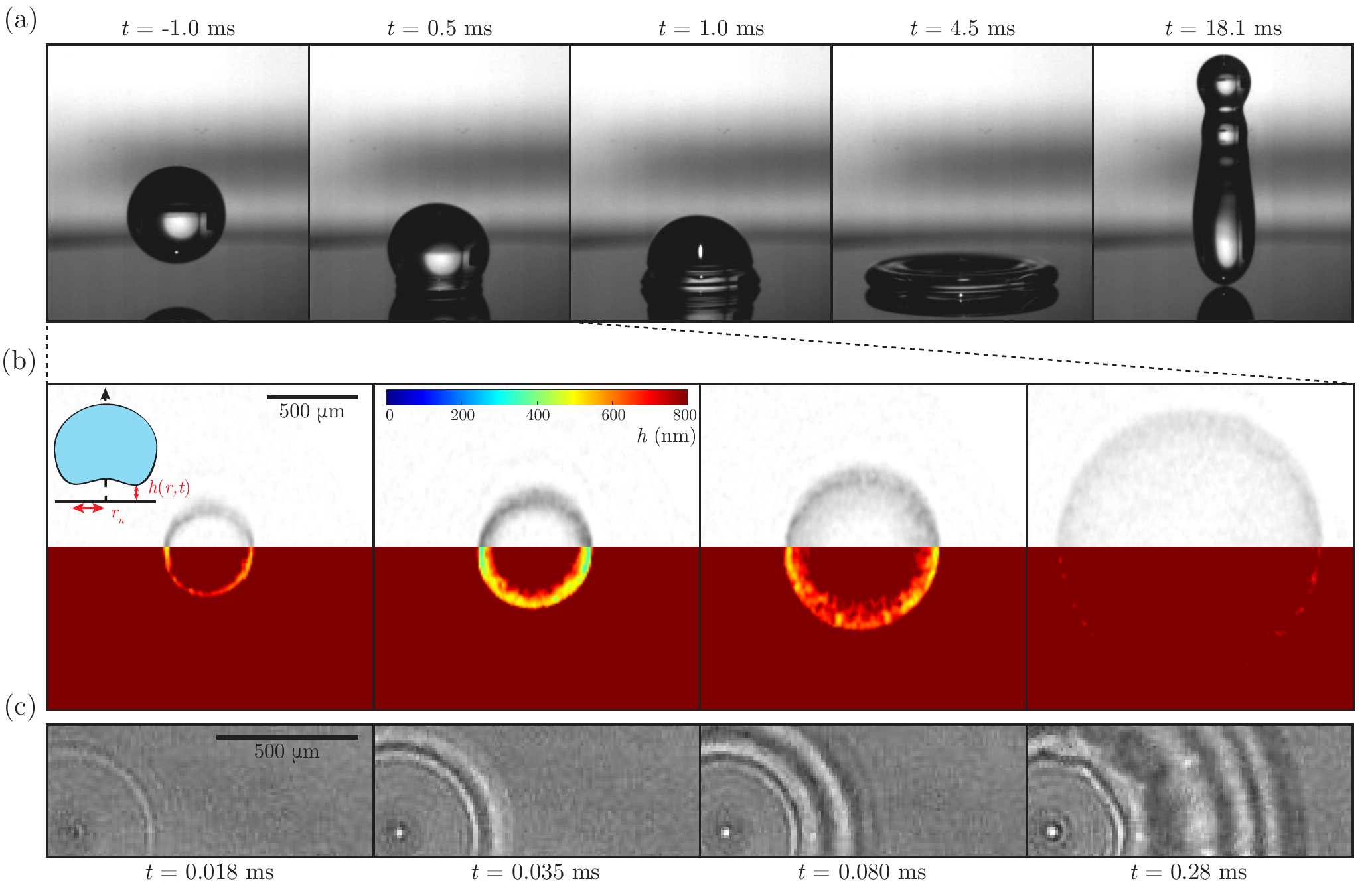}
  \caption{(a) Side view snapshots of the impact of an ethanol drop ($R = 1.1$ mm, $U = 0.5$ m/s, and $\Wen = 9.9$) on a substrate heated at $T_s = 164$$^\circ$C. The liquid detaches after 18.1 ms. Note that the side view is recorded at a slight angle from the horizontal. (b) Short-time TIR snapshots taken during the impact pictured in (a). The origin of time ($t=0$) is chosen as the first frame where the liquid enters within the evanescent lengthscale. The original grayscale image and reconstructed height field with a cutoff height of 800 nm are shown. (c) Synchronized RI snapshots showing approximately one fourth of the drop's bottom interface. Videos (S1 - S2) are available in the supplementary material.}
\label{fig1}
\end{figure}

In figure \ref{fig1}b and c, we present synchronized TIR (figure \ref{fig1}b) and RI (figure \ref{fig1}c) images recorded during the first instants of the impact presented in figure \ref{fig1}a. The TIR snapshots combine the original grayscale image and the calculated height field colormap.
The liquid first enters within the evanescent lengthscale as a faint ring ($t = 0.018$ ms).
This ring expands as its height decreases, indicating that the liquid comes closer to the substrate as it starts spreading ($t = 0.035$ ms).
The full film profile can be inferred by combining the information extracted from TIR with the axisymmetric interference patterns observed in RI images. The brightest fringes are localized at the same radial position as the TIR ring evidencing the region closest to the substrate which is called the neck.
The liquid-air interface displays a dimple shape, sketched in the inset, as already observed in both isothermal and superheated conditions \citep{shirota2016}.
Eventually, the whole bottom interface of the drop moves away from the substrate and escapes from the TIR measurement range ($t = 0.28$ ms) while we continue to observe the spreading liquid as the RI fringe pattern remains visible.

\begin{figure}
  \centering
  \includegraphics[width=1\textwidth]{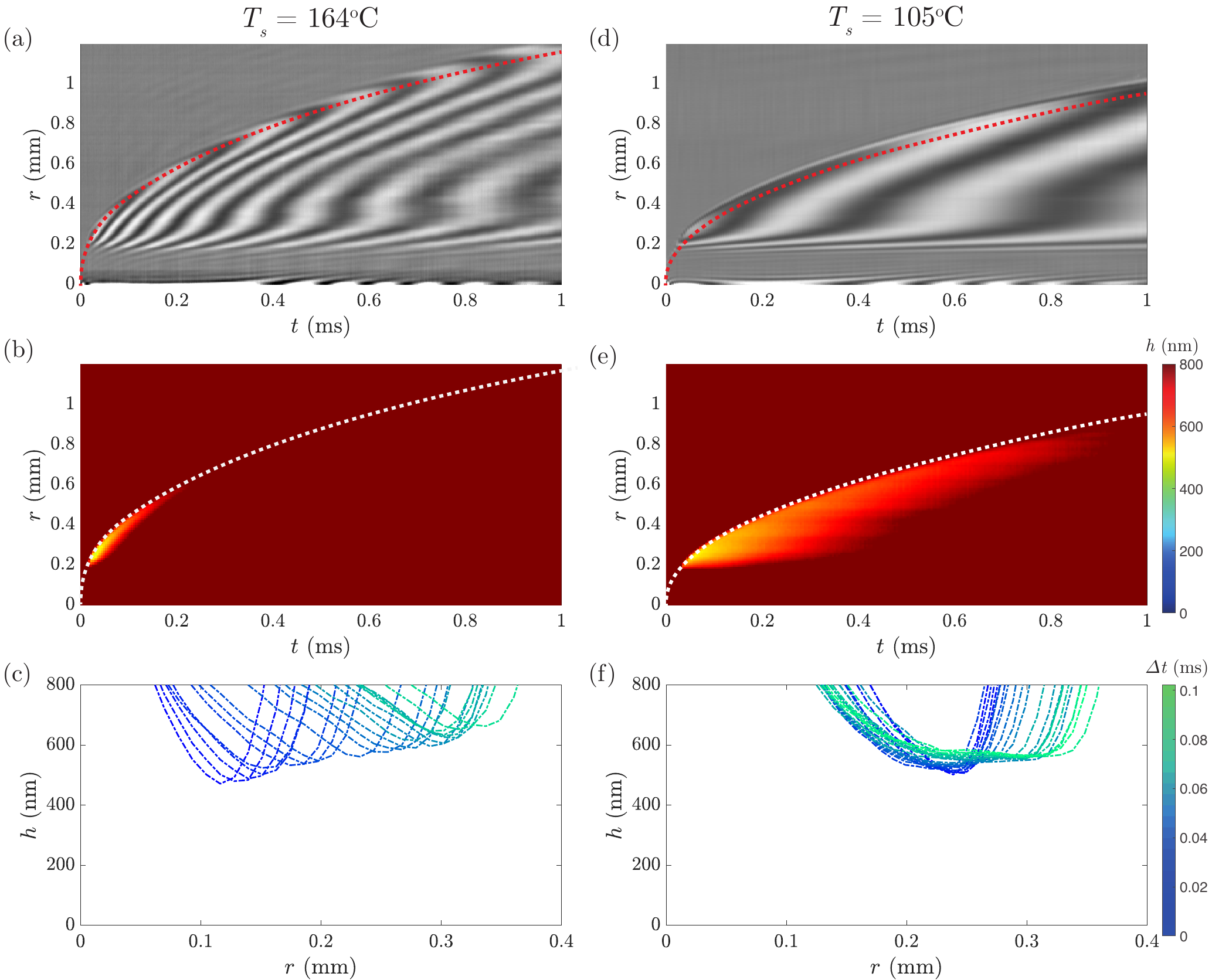}
  \caption{(a)-(b) Space-time plots of the short-time film dynamics obtained by azimuthally averaging the RI and TIR snapshots for the impact shown in figure \ref{fig1}a ($R = 1.1$ mm, $U = 0.5$ m/s, $\Wen = 9.9$ and $T_s = 164$$^\circ$C). The dashed lines in (a)-(b) and (d)-(e) indicate the position of the neck as determined from the RI space-time plots. (c) Successive liquid-gas profiles extracted from (b). The color code stands for the time difference from the instant at which the minimum thickness is reached. (d)-(e) Space-time plots obtained for similar impact parameters as in (a)-(b) ($R = 1.0$ mm, $U = 0.48$ m/s, and $\Wen = 8.3$) but with $T_s = 105$$^\circ$C. (f) Successive height profiles extracted from e. The color code is the same as in (c). Video (S3) is available in the supplementary material.}
\label{fig2}
\end{figure}

Combining TIR and RI, we are able to visualize the dynamics of the whole gas layer and to compare them with their isothermal counterpart.
We azimuthally average each RI and TIR snapshot and stack the one-dimensional information obtained at each instant in space-time two-dimensional graphs (figures \ref{fig2}a and \ref{fig2}b).
From this representation, we are able to distinguish three phases in the early evolution of the squeezed film:
\begin{itemize}
  \item The initial approach during which the liquid-gas interface first deforms, creating the central dimple bordered by a region of high curvature (see the first profile of the liquid-gas interface, $\Delta t = 0$ ms, in figure \ref{fig2}c). This region, which is called the neck region, moves down and outwards until the minimum thickness is reached at the neck, here at $t = 0.031$ ms, marking the end of this first phase. During this initial phase, the motion of the liquid-gas interface is qualitatively similar to that reported in the absence of substrate heating \citep{mandre2009,hicks2010,mani2010,bouwhuis2012}.
  \item A second phase, for $0.031$ ms $<t<0.4$ ms, in which the center of the dimple and the neck have markedly different vertical motion. The thickness at the center ($r=0$) is constant while it increases at the neck, as shown by the TIR data (figure \ref{fig2}b) and by the multiple crossings between the dashed red line, materializing the neck location, and iso-height lines (\emph{i.e} fringes) in figure \ref{fig2}a. Here, the radial and updwards motion of the neck allows to distinguish the superheated from the isothermal case where the neck has a fixed radial position and height \citep{kolinski2012,kolinski2014}.
  \item In the final phase ($t>0.4$ ms), both the thickness at the center and at the neck increase as the liquid keeps spreading. This global film thickness increase, which occurs at different instants at the neck and at the dimple, further differentiates the hot and the cold case and is characteristic of the influence of vapor generation.
\end{itemize}

\subsubsection{Recovering the isothermal behavior}
We observe the transition from the superheated to the isothermal behavior by performing impacts on substrates heated just above the boiling temperature of ethanol.
For similar impact parameters as in figure \ref{fig1}a ($U = 0.48$ m/s, $\Wen = 8.3$) but with $T_s = 105$$^\circ$C, the liquid-gas interface transiently forms a wide flat region that is closer to the substrate (figure \ref{fig2}e), while the dimple shape remains frozen as shown by the horizontal fringes in figure \ref{fig2}d.
The collapse of the successive height profiles at the neck region plotted in figure \ref{fig2}f illustrates the transition from a neck sweeping motion (figure \ref{fig2}c) to a quasi-invariant film shape with decreasing superheat.
The growth of the thickness at the neck and, later, at the center indicates that vapor is still generated and distinguishes this case from the isothermal impact.

We now seek to quantitatively characterize the effect of the superheat on the gas layer. To do so, we describe the neck motion and measure the film thickness at the neck and at the center, when there is no liquid-solid contact at short-time, for different superheats and impact velocities.

\subsection{Neck dynamics}
\begin{figure}
  \centering
  \includegraphics[width=1\textwidth]{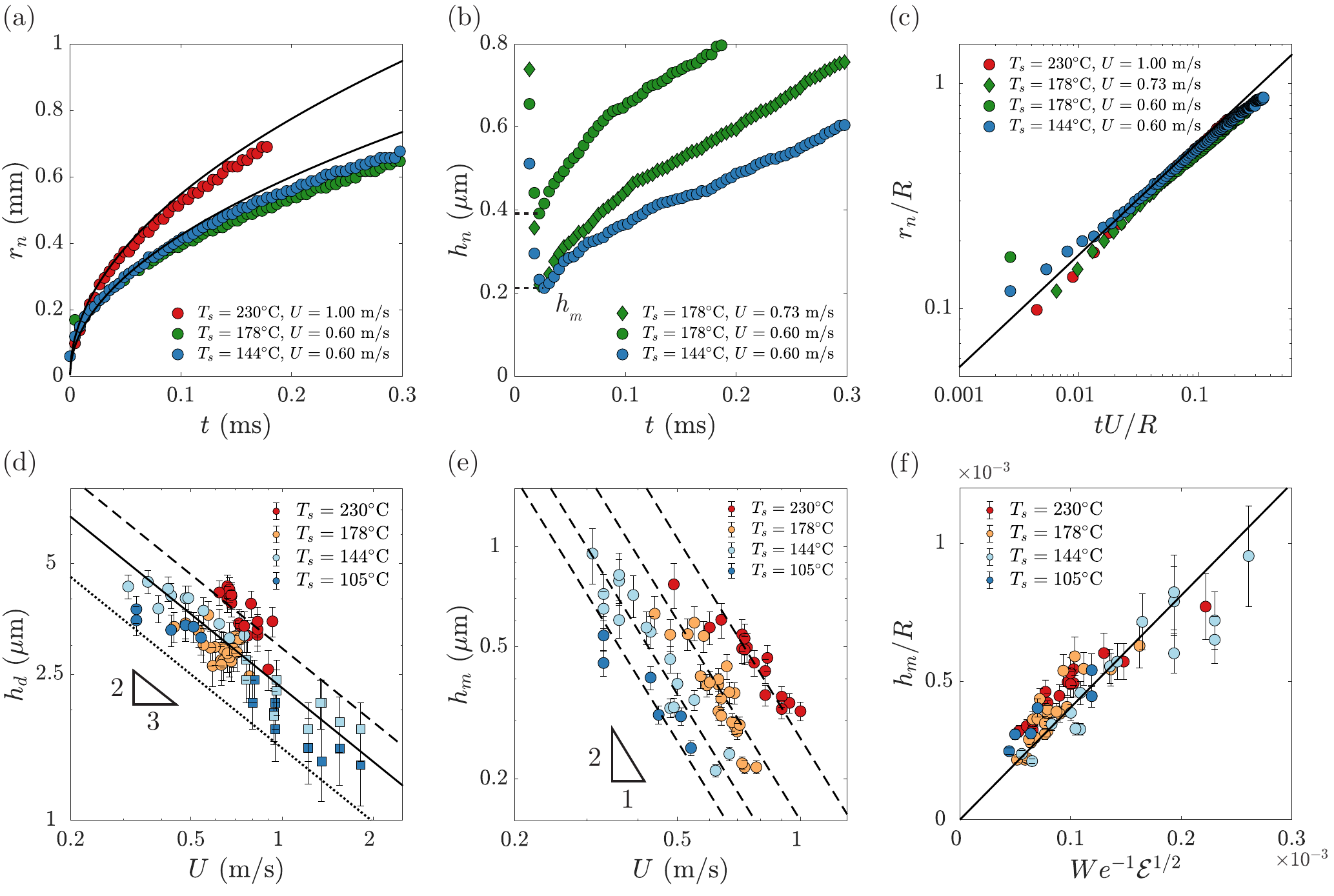}
  \caption{(a) Time evolution of the azimuthally averaged neck radius $r_n(t)$ for varying impact velocities $U$ and substrate temperatures $T_s$. The data are extracted from TIR images such as figure \ref{fig2}b. The solid lines represent predictions from the Wagner theory, $r_n(t) = \sqrt{3URt}$. (b) Azimuthally averaged height at the neck $h_n(t)$ at short-time. We denote $h_m$ the azimuthally averaged minimum film thickness. (c) Normalized azimuthally averaged neck radius $r_n/R$ as a function of the dimensionless time $tU/R$. The solid line shows the Wagner prediction. (d) Central dimple height $h_d$ plotted as a function of $U$ for the impact of drops for various $T_s$. Note that $h_d$ has a non-monotonic dependance with $T_s$. The solid line corresponds to the scaling $h_d/R \sim \Stn^{-2/3}$, taking the viscosity of the gas, $\eta_{g}$, as the viscosity of air at 20$^\circ$C. The dashed and dotted lines correspond to two limiting choices for the gas viscosity, that of air at 230$^\circ$C, and that of ethanol vapor at $T_b$, respectively. The square symbols are obtained in cases in which the liquid touches the solid. (e) Minimum film thickness $h_m$ as a function of the impact speed $U$ for various $T_s$. The dashed lines are guides to the eye with a slope of $-2$. (f) Normalized minimum film thickness $h_m/R$ as a function of the prediction of equation (\ref{hmscaling}), the data are consistent with this prediction and a fit gives a prefactor of $4.0 \pm 0.2$ in equation (\ref{hmscaling}).}
\label{fig3}
\end{figure}

We track the azimuthally averaged neck radius $r_n(t)$ (figure \ref{fig3}a) and distance to the substrate (\emph{i.e} neck height) $h_n(t) = h(r_n(t),t)$ (figure \ref{fig3}b) for varying impact velocities and substrate temperatures at short times.
As the neck spreads, its height decreases, with a velocity $d h_n/dt$ on the order of the impact velocity, until it reaches a minimum value $h_m$ that corresponds to the global minimum of the gas layer thickness. Later, the neck height grows at a slower rate.
The evolution of $r_n(t)$ seems to be only affected by the impact velocity.
In contrast, the neck height $h_n(t)$ is strongly influenced also by the substrate temperature $T_s$.
This behavior suggests that the Wagner prediction, $r_n(t) = \sqrt{3URt}$, shown to be accurate for the radius of the liquid-solid contact by \citet{riboux2014,gordillo2019} on cold surfaces and by \citet{shirota2016} on superheated substrates, could also be relevant in the levitated regime.
We indeed observe a good agreement of that prediction with the data by plotting the normalized neck radius $r_n/R$ as a function of the dimensionless time $tU/R$ for three different impact velocities and superheats (figure \ref{fig3}c). The data collapse, in quantitative agreement with Wagner's prediction represented by the solid line.
We conclude that vapor generation strongly affects the vertical motion of the liquid-gas interface at the neck, but has negligible influence on its horizontal dynamics.

\subsection{Thickness at the dimple and the neck}
We now focus on the effect of $T_s$ on the thickness of the gas film.
In figure \ref{fig3}d, we show the dimple height $h_d$ measured at the same instant as the minimum thickness $h_m$ is reached and, for the two lowest substrate temperatures, also at the instant when contact occurs at the neck (square markers), enabling us to probe velocities up to 2 m/s. The dimple height is on the order of 1 to 5 $\mu$m, and it decreases with increasing impact velocity.
We compare our measurements to the scaling relation for $h_d$ derived in the absence of heating, $h_d/R \sim \Stn^{-2/3}$, where $\Stn = \rho_lRU/\eta_g$ is the Stokes number \citep{mandre2009,mani2010,bouwhuis2012}.
The data are compatible with the isothermal scaling: for a fixed superheat $\Delta T$ and drop radius $R$, the dimple height decrease with increasing impact velocity $U$ is consistent with a power-law behavior with an exponent $-2/3$. Yet, we observe a weak influence of the substrate temperature on $h_d$ that has a non-monotonic dependance with increasing superheat.

In figure \ref{fig3}e, we show the minimum distance $h_m$ separating the drop from the substrate as a function of the control parameters $U$ and $T_s$.
This minimum distance is on the order of a few hundred nanometers, and strongly increases with the superheat, in contrast to the weak non-monotonic influence observed for $h_d$. For example, increasing $T_s$ from 178 to 230$^\circ$C leads to a doubling of the thickness at the neck for $U = 0.7$ m/s.
For a fixed substrate temperature and drop radius, the data suggest a power-law decrease of $h_m$ with the impact velocity $U$ with an exponent $-2 \pm 0.2$. This exponent is close to that predicted for the minimum thickness at the neck in the isothermal case $-20/9$ \citep{mandre2009,mani2010}, but the strong influence of $\Delta T$ evidences the role of evaporation and distinguishes this case from an impact on a substrate at ambient temperature .

\subsection{Model for the initial approach}
\label{approachmodel}
We now seek to model the initial approach, \emph{i.e}, the evolution of the gas film until the minimum thickness is reached at the neck.
Typically, this phase lasts until $tU/R < 0.01$, a time much shorter than that associated with the purely inertial spreading that is valid up to $tU/R < 0.1$.
We build on the work of \citet{mani2010}, who treated the case of isothermal impacts, and extend it to take into account the effect of vapor production.
Indeed, our measurements of the dimple thickness in superheated conditions show the same scaling relation with the Stokes number as for isothermal impacts, indicating that liquid inertia and viscous drainage are still the relevant mechanisms when the substrate is heated. Yet, the thickness at the neck strongly depends on the superheat, suggesting that, to describe the movement of the liquid-gas interface, one also needs to capture the influence of vapor generation.

For completeness, we reproduce the equations of motion for the liquid and for the gas film in the incompressible regime as already derived by \citet{mani2010}.
Following them, we consider a two-dimensional geometry and make further simplifications as in the isothermal case: in the liquid, we neglect viscosity and non-linear inertia owing to, respectively, the large Reynolds number ($\Rey = \rho_l R U/\eta_l = 330$ for the impact of a millimeter-sized drop at $U = 0.5$ m/s) and the absence of velocity gradients within the drop before it interacts with the substrate.
We also neglect the influence of surface tension, as for a typical impact the Weber number, $\Wen \approx 10$, is considerably larger than unity.
With these assumptions, we have potential flow in the liquid and the equation of motion reads:
\begin{equation}
  \rho_l \frac{\partial\mathbf{u}}{\partial t} + \mathbf{\nabla} p_l = 0, \qquad \nabla \cdot \mathbf{u} = 0
  \label{liquid1}
\end{equation}
where $\mathbf{u} = (u,v)$ are the liquid velocity components in the $x$ (replacing $r$ in this 2D model) and $z$ directions, respectively, and $p_l$ is the liquid pressure.
Projecting equation (\ref{liquid1}) in the vertical direction and using the kinematic boundary condition at the liquid-gas interface $\partial h/\partial t = v - u \partial h/\partial x$, we obtain at $z=0$:
\begin{equation}
  \rho_l \frac{\partial^2 h }{\partial t^2} + \frac{\partial p_l}{\partial z} = - \rho_l \frac{\partial}{\partial t}\left(u\frac{\partial h}{\partial x}\right)
  \label{liquideq0}
\end{equation}
where, in the spirit of the boundary layer approximation \citep{prandtl1904}, the term on the right-hand side can be neglected. Then, the vertical pressure gradient at the interface can be expressed as the Hilbert transform of the horizontal pressure gradient by taking advantage of the two-dimensional nature of the problem and the harmonic pressure field \citep{smith2003}:
\begin{equation}
  \rho_l \frac{\partial^2 h }{\partial t^2} -\mathcal{H}\left[\frac{\partial p_l}{\partial x}\right] = 0.
  \label{liquideq}
\end{equation}
We now come to the description of the gas flow between the drop and the substrate. Here, the viscous lubrication approximation is justified as the gas film is thin ($h \ll R$) and the typical value of the Reynolds number is low in the gas phase ($\Rey = \rho_g h_d U/\eta_g$ = 0.05 for air at 20$^\circ$C). This approximation is valid when considering heating as the value of the Reynolds number is not significantly altered by the presence of ethanol vapor ($\Rey = \rho_v h_d U/\eta_v$ = 0.15 for ethanol vapor at $T_b$), nor by the dependance of the gas viscosity and density on temperature. With these assumptions, we obtain the lubrication equation:
\begin{equation}
  \frac{\partial h}{\partial t} - \frac{\xi}{12 \eta_g}\frac{\partial}{\partial x}
  \left(h^3 \frac{\partial p_g}{\partial x}\right) = 0,
  \label{lubeqnovap}
\end{equation}
where $\xi$ is a numerical coefficient which accounts for the choice of boundary condition at the liquid-gas interface. If the no-slip condition holds at the interface, $\xi = 1$. In constrast, if a zero tangential stress condition is chosen, $\xi =4$ \citep{petit2012}.

We now introduce an additional term in the lubrication equation to take into account the effect of evaporation.
We consider that heat is transferred through the gas layer by conduction, similarly as in the static Leidenfrost situation, as this mechanism acts on a relevant timescale for impact, $h^2/\kappa_g \approx 0.1$ $\mu$s, much shorter than the initial approach time.
The energy input from the heated solid is used to heat liquid at the bottom interface from the ambient temperature $T_a$ to the boiling temperature $T_b$, creating a thermal boundary layer with thickness $\sqrt{\kappa_l t}$, and to vaporize it.
The Jakob number $Ja = C_{p,l}(T_b - T_a)/\mathcal{L}$, which compares the sensible heat with the latent heat $\mathcal{L}$, is $0.16$ so that we consider the energetic cost coming from the latent heat to be dominant.
With this assumption, which overestimates the vapor production, the evaporation rate per unit area, $e$, is given by balancing the heat flux $k_g\Delta T/h$ with the released latent heat $\mathcal{L}e$, yielding $e = k_g\Delta T/(\mathcal{L}h)$.
We thus obtain a modified lubrication equation:
\begin{equation}
  \frac{\partial h}{\partial t} - \frac{\xi}{12 \eta_g}\frac{\partial}{\partial x}
  \left(h^3 \frac{\partial p_g}{\partial x}\right) = \frac{1}{\rho_g}\frac{k_g \Delta T}{\mathcal{L}h}
  \label{lubeq}
\end{equation}
where the term on the right hand side takes into account evaporation \citep{biance2003,sobac2014}.
\subsubsection{Dimple height}
Having obtained the governing equations for the liquid and the gas, we recall the dimple height scaling derived for isothermal impacts \citep{mani2010,bouwhuis2012}. It is obtained by balancing liquid inertia with viscous drainage in the squeezed gas film.
The pressure in the liquid can be estimated from equation (\ref{liquideq0}) or (\ref{liquideq}):
\begin{equation}
   p_l/L \sim \rho_l U^2/h_d
\end{equation}
where $L \sim \sqrt{h_dR}$ is the radial extent of the dimple computed as the radius of a spherical cap with height $h_d$.
The gas pressure is deduced from the two-dimensional incompressible lubrication equation (\ref{lubeqnovap}):
\begin{equation}
  p_gh^3/\eta_gL^2 \sim U.
\end{equation}
The dimple height is set when the liquid-gas interface first deforms, that is when the pressure in the liquid and in the gas become comparable:
\begin{equation}
  h_d \sim R\left(\frac{\eta_g}{\rho_lRU}\right)^{2/3} \sim R \Stn^{-2/3}
  \label{hd}
\end{equation}
where $\Stn = \rho_lRU/\eta_g$ is the Stokes number.
The solid line (figure \ref{fig3}d) represents equation (\ref{hd}) with $\eta_g$ taken as the viscosity of air at 20$^\circ$C and a prefactor of 2.8 extracted from the results of \citet{bouwhuis2012} for the impact of ethanol drops on a substrate at room temperature. The isothermal scaling gives a correct order of magnitude of $h_d$ and recovers the velocity dependance in superheated conditions, confirming that the viscous drainage of the gas layer dominates vapor generation as suggested by the weak influence of substrate temperature on the dimple height in our experiments.

We propose to explain the observed influence of $T_s$ also through equation (\ref{hd}), although it does not explicitly involve temperature. Indeed, $\eta_g$ depends on temperature in two ways.
(i) The viscosity of the air squeezed between the drop and the surface increases with $T_s$, and
(ii) as ethanol vapor is generated, the squeezed layer becomes a - possibly non-homogeneous - mixture of warm air and ethanol vapor, whose viscosity is lower than that of air at the same temperature.
The interplay between these two antagonistic effects could be the cause of the non-monotic behavior of $h_d$ with $T_s$.
In figure \ref{fig3}d, we plot the scalings associated with each effect by taking $\eta_g$ as the viscosity of air at 230$^\circ$C (dashed line) and as the viscosity of ethanol vapor at the boiling point (dotted line): all measurements lie in between the two bounds.

\subsubsection{Neck thickness}
We next focus on the subsequent formation of the neck. Here, we extend the calculation of \citet{mani2010} to incorporate the effect of evaporation, as our experimental observations show a strong influence of the substrate temperature on the thickness at the neck.
We non-dimensionalize the governing equations (\ref{liquideq}) and (\ref{lubeq}) with the scales involved in the dimple formation, namely with the transformations:
\begin{equation}
  h = R \Stn^{-2/3}\tilde{h}, \quad x = R \Stn^{-1/3}\tilde{x}, \quad t = \frac{R \Stn^{-2/3}}{U}\tilde{t}, \quad p_l = \frac{\eta_g U}{R \Stn^{-4/3}}\tilde{p}_l, \quad p_g = P_0 \tilde{p}_g
\end{equation}
where $P_0$ is the atmospheric pressure.
Then, the equations of motion of the liquid and gas respectively become:
\begin{equation}
  \frac{\partial^2 \tilde{h}}{\partial \tilde{t}^2} - \varepsilon \mathcal{H}\left[ \frac{\partial \tilde{p}_l}{\partial \tilde{x}}\right] = 0,
  \label{liquideqscaled}
\end{equation}
\begin{equation}
  \frac{\partial \tilde{h}}{\partial \tilde{t}} - \delta \frac{\xi}{12}\frac{\partial}{\partial \tilde{x}}\left(\tilde{h}^3\frac{\partial \tilde{p}_g}{\partial \tilde{x}}\right) = \frac{\mathcal{E}\Stn^{5/3}}{\Wen \tilde{h}}.
  \label{lubeqscaled}
\end{equation}
Here we have introduced two extra dimensionless quantities:
\begin{itemize}
\item the ratio of the atmospheric pressure to the pressure build-up below the drop,
\begin{equation}
    \delta = \frac{P_0 RSt^{4/3}}{\eta_gU},
\end{equation}
\item and the evaporation number $\mathcal{E}$ defined by \citet{sobac2014} in the study of static Leidenfrost drops,
\begin{equation}
  \mathcal{E} = \frac{\eta_g k_g \Delta T}{\gamma \rho_g R \mathcal{L}}.
  \label{evapn}
\end{equation}
\end{itemize}
The pressure buildup in the gas film below the drop is a small correction to the atmospheric pressure. We thus consider the limit $\delta \gg 1$, and assume the following pressure expansion $\tilde{p}_g = 1 + \tilde{p}/\delta$ which we introduce in the governing equation:
\begin{equation}
  \frac{\partial^2 \tilde{h}}{\partial \tilde{t}^2} - \mathcal{H}\left[ \frac{\partial \tilde{p}}{\partial \tilde{x}}\right] = 0, \qquad
  \frac{\partial \tilde{h}}{\partial \tilde{t}} - \frac{\xi}{12}\frac{\partial}{\partial \tilde{x}}\left(\tilde{h}^3\frac{\partial \tilde{p}}{\partial \tilde{x}}\right) = \frac{\mathcal{E}\Stn^{5/3}}{\Wen \tilde{h}},
  \label{goveq}
\end{equation}
where we used the pressure continuity at the interface (\emph{i.e} $p_g = p_l$).
As the motion of the liquid-gas interface is similar in the isothermal and the superheated cases, we adopt the same self-similar ansatz as in \citet{mani2010} and construct a self-similar solution for the behavior in the neck region. The height and the pressure are given by:
\begin{equation}
  \tilde{h}(\tilde{x},\tilde{t}) = \tilde{h}_n(\tilde{t})H(\Theta), \qquad \tilde{p}(\tilde{x},\tilde{t}) = \tilde{p}_n(\tilde{t})\Pi(\Theta)
\end{equation}
where $\Theta(\tilde{x},\tilde{t}) = (\tilde{x}-\tilde{x}_n(\tilde{t}))/\tilde{\ell}(\tilde{t})$ is the self-similarity variable, $\tilde{x}_n(\tilde{t})$ the neck's radial coordinate, $\tilde{p}_n(\tilde{t})$ the pressure at $\tilde{x} = \tilde{x}_n(\tilde{t})$, and $\tilde{\ell}(\tilde{t})$ the horizontal length scale associated with the high-curvature region, \emph{i.e} the neck region.
The time derivatives of the height or pressure field have three contributions coming respectively from the height's temporal variation, the change of horizontal extent, and the radial motion of the neck:
\begin{equation}
  \frac{\partial \tilde{h}}{\partial \tilde{t}} = \frac{d \tilde{h}_n}{d\tilde{t}}H - \frac{\tilde{h}_n \Theta}{\tilde{\ell}} \frac{d \tilde{\ell}}{d\tilde{t}}\frac{d H}{d \Theta} - \frac{\tilde{h}_n}{\tilde{\ell}}\frac{d \tilde{x}_n}{d\tilde{t}}\frac{d H}{d \Theta}.
\end{equation}
\citet{mani2010} hypothesized that the advection term is dominant and proposed that the neck region has a wave-like behavior: $d/d\tilde{t} \approx \tilde{c} d/d\tilde{x}$ where $\tilde{c} = d \tilde{x}_n/d\tilde{t}$ is the neck radial velocity that is considered to be constant during the initial approach. This hypothesis seems to be at odds with our description of the neck dynamics using Wagner theory where $x_n$ evolves as $\sqrt{t}$.
This apparent contradiction can be resolved by comparing the timescales associated with the initial approach phase and the motion captured by Wagner's theory. Typically, the minimum thickness is reached for $tU/R < 0.01$  while Wagner's theory is valid up to dimensionless times on the order of $0.1$. This separation of timescales validates the use of a linear approximation for the neck position during the initial approach.
Introducing the self-similar fields in equations (\ref{goveq}) and using the wave like nature of the solution we obtain:
\begin{equation}
  \tilde{h}_n\frac{\tilde{c}^2}{\tilde{\ell}^2}\frac{dH}{d\Theta} = \frac{\tilde{p}_n}{\tilde{\ell}}\mathcal{H}\left[\frac{d\Pi}{d\Theta}\right],
  \label{selfsimliquid}
\end{equation}
\begin{equation}
  \tilde{h}_n\frac{\tilde{c}}{\tilde{\ell}}\frac{dH}{d\Theta} - \frac{\xi}{12}\frac{\tilde{h}_n^3\tilde{p}_n}{\tilde{\ell}^2}\frac{d}{d\Theta}\left(H^3\frac{d\Pi}{d\Theta}\right) = \frac{\mathcal{E}\Stn^{5/3}}{\Wen\tilde{h}_n H}.
  \label{selfsimgas}
\end{equation}
This set of equations is identical to that obtained by \citet{mani2010} in the isothermal case except for the additional term on the right hand side of equation (\ref{selfsimgas}) that incorporates the influence of evaporation in the viscous lubrication flow.

If the liquid's latent heat tends towards infinity or no superheat is applied, the evaporation number $\mathcal{E}$ tends towards zero and the terms on the left hand side balance as in the isothermal case (see \citet{mani2010} and appendix \ref{appendix_hmin}).
On the contrary, if evaporation plays a dominant role as suggested by the strong influence of superheat on the minimum thickness at the neck, the evaporation term balances the downwards motion of the interface (the first term in equation (\ref{selfsimgas})).
Equation (\ref{selfsimgas}) then gives a relationship between the horizontal and the vertical lengthscales:
\begin{equation}
  \tilde{\ell} \sim \tilde{c} \tilde{h}_n^2 \mathcal{E}^{-1} \Stn^{-5/3} \Wen,
\end{equation}
that we combine with the scaling extracted from equation (\ref{selfsimliquid}), $\tilde{p}_n \sim \tilde{h}_n \tilde{c}^2/\tilde{\ell}$, to obtain a relation between the pressure and height at the neck:
\begin{equation}
  \tilde{p}_n \sim \tilde{c}\tilde{h}_n^{-1}\mathcal{E}\Stn^{5/3}\Wen^{-1}.
\end{equation}
As $h_n$ decreases, the initially neglected Laplace pressure at the neck $\gamma h_n/\ell^2$, that scales as $h_n^{-3}$, diverges quicker than the gas pressure at the neck $p_n$, that evolves as $h_n^{-1}$. The hypothesis to neglect surface tension is no longer valid as the drop approaches the solid, similarly as for isothermal impacts.
The balance between $p_n$ and the Laplace pressure sets the minimum film thickness $h_m$:
\begin{equation}
  \frac{h_m}{R} \sim \Wen^{-1}\mathcal{E}^{1/2}.
  \label{hmscaling}
\end{equation}
Equation (\ref{hmscaling}) predicts a power-law decrease of the minimum thickness as a function of impact velocity $U$ with an exponent $-2$ when fixing the drop radius and superheat, consistent with the experimental findings displayed in figure \ref{fig3}e, where the dashed lines are guides to the eye with a slope of $-2$.
This power-law decrease, $h_m \propto U^{-2}$, is close to that numerically predicted in the isothermal case where the minimum thickness follows a power-law with exponent $-20/9$ \citep{mani2010}. In appendix \ref{appendix_hmin}, we derive this isothermal scaling and show that it does not allow to capture the effect of substrate temperature contrary to equation (\ref{hmscaling}) which explicitly involves superheat.
Testing the effect of the substrate temperature $T_s$ requires to take into account the temperature-dependant values of the gas viscosity, density and thermal conductivity as well as the value of the liquid surface tension.
The viscous effects are introduced by the gas drainage associated to dimple formation. We thus use the temperature-dependant gas viscosity extracted from our measurements of the dimple height.
The gas density and thermal conductivity, on the contrary, are related to vapor generation at the neck.
Given the conduction timescale $h^2/\kappa_g\approx 0.1$ $\mu$s, we consider steady heat transfer in the lubrication layer. The temperature profile is linear between the substrate temperature $T_s$ and the temperature at the liquid-gas interface where evaporation occurs, that is $T_b$.
We then evaluate the gas density $\rho_g$ and thermal conductivity $k_g$ at $(T_s+T_b)/2$, assuming that the gas phase in the neck region is constituted of ethanol vapor only, and taking the surface tension $\gamma$ at temperature $T_b$.
In figure \ref{fig3}f, we compare the normalized minimum thickness $h_m/R$ to the prediction of equation (\ref{hmscaling}). The data for different superheat and impact velocity collapse on a line with prefactor $4.0\pm 0.2$, a satisfactory agreement considering the complexity of the system.

\section{Transition from levitation to contact}
As the minimum film thickness decreases with the impact velocity, one expects that for large enough impact velocities the liquid wets the solid at short times; an event that indeed occurs.
In this section, we report our observations of the collapse of the gas film and discuss the transition from levitation to contact.
We first map this transition in the parameter space spanned by surface temperature and impact velocity (figure \ref{fig4}a). We report three distinct types of breakdowns, namely a short-time contact (filled red diamonds), a late-time contact (filled orange dots), and a contact induced by oscillations of the gas film (open red diamonds).
We now discuss the phenomenology of these contacts, starting from large velocities in figure \ref{fig4}a, and then going to lower and lower ones.

\subsection{Phenomenology}
\subsubsection{Short-time contact}
\begin{figure}
  \centering
  \includegraphics[width=1\textwidth]{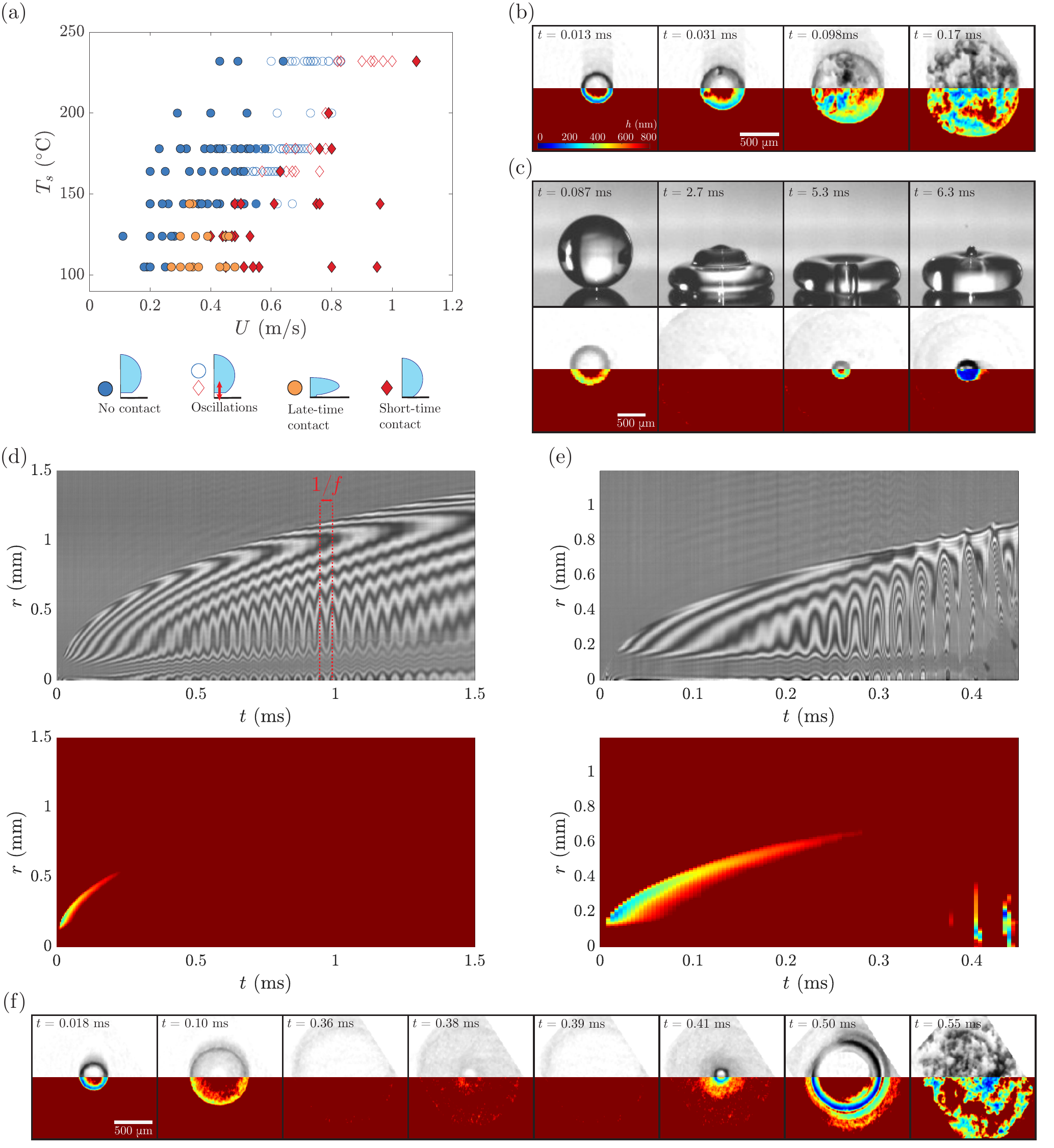}
  \caption{(a) Phase diagram of the levitation and contact regimes for ethanol drops in the parameter space spanned by impact velocity $U$ and substrate temperature $T_s$. (b) TIR snapshots for an impact with $U = 0.76$ m/s, $R = 1.0$ mm (\emph{i.e} $\Wen = 21$), and $T_s = 178$$^\circ$C. Contact occurs at short-time and we observe a partial wetting pattern ($t = 0.17$ ms) characteristic of transition or contact boiling. (c) Synchronized side view and TIR images ($R = 1.0$ mm, $U = 0.33$ m/s, $\Wen = 3.9$, and $T_s = 105$$^\circ$C) showing late-time contact induced by the jet formed during the retraction process. (d) Space-time RI and TIR plots for an impact with $U = 0.68$ m/s, $R = 1.0$ mm (\emph{i.e} $\Wen = 17$), and $T_s = 178$$^\circ$C. Axisymmetric oscillations of the gas film, with frequency $f = 27 \pm 3$ kHz, appear as the central film thickness starts to increase. Their amplitudes eventually decrease and no contact occurs during the whole rebound. Oscillations are not visible using TIR imaging. (e) Space-time plots extracted from RI and TIR for an impact with $U = 0.78$ m/s, $R = 1.0$ mm (\emph{i.e} $\Wen = 22$), and $T_s = 178$$^\circ$C. The oscillations, initially only visible using RI, grow and appear within the evanescent lengthscale as the interference pattern disappears. (f) TIR snapshots for the same impact as in (e), showing the growth of oscillations and the creation of a circular contact ($t = 0.5$ ms). Videos (S4 - S7) are available in the supplementary material.}
\label{fig4}
\end{figure}

We begin by describing the prevalent mode of contact that occurs for all the probed substrate temperatures. Short-time contact is characterized by the occurrence of localized liquid-solid contact at the neck for $t \ll \tau_i$ (figure \ref{fig4}b, $t= 0.031$ ms).
After the contact nucleates, we observe partial wetting patterns typical of transition or contact boiling ($t = 0.098$ and $t= 0.17$ ms), as previously described by \citet{shirota2016}.
The impact velocity $U$ above which short-time contact is observed increases with the substrate temperature $T_s$ (filled red diamonds, figure \ref{fig4}a), as one would expect. As the impact velocity increases, the gas film thins and eventually contact occurs below a critical value given by defects present on the surface or by the attractive surface forces.
For moderate impact velocities ($U < 0.8$ m/s), we have sufficient space and time resolution to determine the film height at the instant before the film collapses (see appendix \ref{appendix_rupture}).
The broad distribution of thicknesses at rupture with an average value of 0.36 $\mu$m, much larger than the root mean square roughness of the sapphire substrate or the range of the attractive surface forces, suggests that defects or surface contamination play a dominant role in triggering liquid-solid contact, in agreement with data obtained for isothermal impact on glass substrates \citep{ruiter2012,kolinski2014_2}.

\subsubsection{Late-time contact}
Next, we discuss the late-time contact mode. It occurs for $T_s < 150$$^\circ$C and at lower velocities than the short-time rupture discussed above (filled orange dots in figure \ref{fig4}a).
In this situation, the drop spreads (figure \ref{fig4}c, $t = 0.087$ and $t = 2.7$ ms) and starts to recoil (figure \ref{fig4}c, $t = 5.3$ ms) while floating on a gas cushion.
Yet, at $t = 5.3$ ms, we observe a decrease of the gas film thickness at the center of the drop that eventually results in wetting (figure \ref{fig4}c, $t = 6.3$ ms).
We deduce from the observation of synchronized side-views and TIR snapshots (figure \ref{fig4}c) that the rupture of the gas film coincides with the ejection of a liquid jet, originating from the collapse of the air cavity during the recoil stage of the rebound.

The formation of air cavities during impact is not specific to superheated substrates: their appearance has been linked to the oscillations of capillary waves excited at impact \citep{renardy2003}.
The collapse of these cavities produces the upwards ejection of a thin jet \citep{bartolo2006}, reminiscent of the bursting of a bubble at a free surface \citep{boulton-stone_1993,deike2018}. Similarly, the ejection of a downwards jet during drop impact has been recently observed using x-ray imaging \citep{lee2020}.
\citet{bartolo2006} showed that the jet ejection velocity can be one order of magnitude larger than the impact velocity, qualitatively explaining why the film rupture can occur at late times for lower impact speeds than at short times.

\subsubsection{Contact induced by oscillations of the gas film}
Finally, for $T_s > 150$$^\circ$C, we observe vertical axisymmetric oscillations of the gas film squeezed below the impacting drop in the transitional regime between levitation and short-time contact (open symbols in figure \ref{fig4}a).
Figure \ref{fig4}d shows a typical space-time RI plot of the initial gas layer dynamics for an impact ($U = 0.68$ m/s, $\Wen = 17$, and $T_s = 178$$^\circ$C) that lies in the transition region.
Oscillations of the liquid-gas interface appear at the center of the gas film after the initial approach phase, when the dimple height starts to grow, and quickly perturb the whole bottom interface as their amplitude increase.
The dynamics of the gas film after the minimum thickness is reached can be decomposed in two contributions that act on separate timescales. (i) A slow evolution, that corresponds to the dynamics of the gas film described in \textsection 3, with a characteristic time on the order of one millisecond and (ii) fast oscillations with a period on the order of a few tens of microseconds.
The displacement of the iso-height lines associated with these fast oscillations occurs in phase: iso-height contours are crossed at the same time at every radial location (figure \ref{fig4}d).
The oscillations of the drop bottom interface thus correspond to a vertical movement of the entire bottom interface and not to wave propagation.
The amplitude of oscillations cannot grow indefinitely, but either saturates and decreases or finally induces film collapse.
The former situation corresponds to the impact pictured in figure \ref{fig4}d, where the drop is in the levitated regime as further evidenced by the space-time TIR data (figure \ref{fig4}d) where the oscillations are invisible.
The latter is illustrated in the space-time representations of figure \ref{fig4}e. As the amplitude increases, the fringe pattern disappears and the oscillating interface comes closer to the substrate and enters within the evanescent lengthscale once again (figure \ref{fig4}e).
The TIR snapshots of figure \ref{fig4}f allow us to visualize the film collapse. Contact occurs along a ring (figure \ref{fig4}f, $t = 0.50$ ms) of a much larger diameter than that of the dimple, and leads to the appearance of partial wetting patterns (figure \ref{fig4}f, $t = 0.55$ ms).

\subsection{A comment on the dynamic Leidenfrost temperature}
Having described the phenomenology of the three collapse modes present in the phase diagram, we now discuss their influence on the dynamic Leidenfrost transition. Figure \ref{fig4}a contains two key pieces of information that give us new insight on the dynamic Leidenfrost temperature.
(i) In the low-velocity limit, the static Leidenfrost temperature, that is approximately 150$^\circ$C for ethanol, does not act as a lower asymptote for the dynamic Leidenfrost temperature. The later can fall below the static value as a consequence of the transient stability of the gas cushion promoted by drainage. Thus, remarkably, dynamic levitation can be less demanding than static levitation.
(ii) The gas film collapse can be driven by multiple mechanisms: namely short-time contact, late-time contact and contact induced by vertical oscillations of the gas film.
Above 150$^\circ$C, the later collapse mechanism occurs for lower impact velocities than short-time contact, demonstrating that the transition towards the dynamic Leidenfrost effect is controlled by the presence of oscillations and not directly linked to a critical thickness reached during the initial approach phase.
Predicting the dynamic Leidenfrost temperature thus requires a better understanding of each collapse mode, especially the oscillation induced contact mode, which we further investigate.

\subsection{Oscillations: a minimal model}
We now focus on understanding the appearance of oscillations in the squeezed gas layer, a feature specific to impacts on superheated surfaces unlike short or late-time contact.
Our observations are reminiscent of the rapid vibrations that appear at the base of soft sublimable solids as contact occurs with a hot substrate during an impact \citep{waitukaitis2017,waitukaitis2018}.
Oscillations can also spontaneously develop, without contact, at the bottom interface of drops in the static Leidenfrost state \citep{bouillant2020} or levitated by a steady airflow \citep{bouwhuis2013}, and are responsible for the subsequent emergence of star shapes \citep{holter1952,brunet2011}.
For both levitation mechanisms, the appearance of vertical oscillations has been linked to the coupling of drop motion and flow in the thin gas film. Although the frequencies reported in these systems are on the order of $100$ Hz, two orders of magnitude smaller than those we measure at impact, our observations can be related to this hydrodynamic mechanism.
(i) Oscillations grow as the dimple height starts to increase, that is when the vapor flow influences the whole liquid-gas interface.
(ii) There is a temperature threshold below which vibrations do not appear, \emph{i.e}, a sufficient gas flow is needed to make the interface unstable.

We thus build an hydrodynamic model that accounts for the motion of the bottom interface after the minimum thickness is reached at the neck by adapting the model of \citet{bouillant2020} to the impact situation.
We reproduce here for completeness the derivation of the equation coupling the drop motion and the lubrication flow created by vapor generation. We apply Newton's second law to the vaporizing drop in the vertical direction.
The momentum variation has two sources: the motion of the drop's center of mass, $m\mathrm{d}^2 z_{cm}/\mathrm{d} t^2$, and the ejection of vapor, $v_{e}\mathrm{d} m/\mathrm{d} t$ where $m(t)$ is the drop mass, $z_{cm}(t)$ the vertical position of its center of mass and $v_{e}(t)$ the vertical ejection velocity of vapor in the reference frame of the drop.
We consider a simplified geometry, that of a gas layer with uniform thickness $h(t)$ and constant radial extent $R$, allowing us to derive the evaporation rate and the lubrication force analytically \citep{biance2003}.
Indeed, we obtain the evaporation rate $\mathrm{d} m/\mathrm{d} t$ by integrating the evaporation rate per unit area $e = - k_v\Delta T/(\mathcal{L} h)$, derived in section \ref{approachmodel}, over the bottom interface of the drop assuming that evaporation predominantly occurs in the gas layer.
The lubrication force is obtained by integrating the pressure $p_g(r,t)$, derived from the lubrication equation (\ref{lubeq}), over the gas layer: $F_L(t) = \int_0^R p_g(r,t) 2\pi r \mathrm{d}r$.
We then write the momentum balance:
\begin{equation}
  m\frac{\mathrm{d}^2 z_{cm}}{\mathrm{d} t^2} + \frac{k_v \Delta T\pi R^2}{\mathcal{L} h} v_{e} = -\frac{6\eta_v\pi R^4}{\xi h^3}\left(\frac{\mathrm{d}h}{\mathrm{d}t} - \frac{1}{\rho_g}\frac{k_v\Delta T}{\mathcal{L} h}\right) -mg,
\end{equation}
and express the ejection velocity $v_e$ as the sum of the absolute ejection velocity $-k_v\Delta T/(\rho_v\mathcal{L}h)$, derived from a mass balance, and of the interface velocity $-\mathrm{d}h/\mathrm{d}t$:
\begin{equation}
  m\frac{\mathrm{d}^2 z_{cm}}{\mathrm{d} t^2} +
  \left(\frac{6\eta_v\pi R^4}{\xi  h^3}-\frac{k_v \Delta T \pi R^2}{ \mathcal{L} h} \right)\frac{\mathrm{d}h}{\mathrm{d}t} -
  \frac{k_v^2 \Delta T^2 \pi R^2}{\rho_v \mathcal{L}^2 h^2} -
  \frac{6\eta_v k_v \Delta T \pi R^4}{\xi\rho_v  \mathcal{L} h^4} - mg = 0.
  \label{momentum1}
\end{equation}

Equation (\ref{momentum1}) is identical to that obtained by \citet{bouillant2020} in the static Leidenfrost situation.
Relating the motion of the bottom interface of the drop with that of its center of mass is the critical step where the static and impact situations differ.
At short times, \emph{i.e} for $t \ll \tau_i = R/U$, the drop is in the kinematic phase. The motion of the drop's center of mass is approximately ballistic: $z_{cm}(t) \approx h(t) + R - Ut$, allowing to recover an equation for the vertical motion of the liquid-gas interface from equation (\ref{momentum1}) which we non-dimensionalize using the following transformations:
\begin{equation}
  h = h_d \hat{h}, \qquad t = \tau \hat{t}
\end{equation}
where we chose the dimple height $h_d$ as length scale as it gives the correct order of magnitude of the film thickness after the initial approach and denote as $\tau$ the characteristic time of the oscillations of the gas film.
Plugging these non-dimensional variables in equation (\ref{momentum1}) allows us to identify the physical phenomena and time scales involved in the gas film dynamics:
\begin{multline}
  \frac{\mathrm{d}^2 \hat{h}}{\mathrm{d} \hat{t}^2} +
  \tau \frac{\beta}{m} \left(\frac{6\pi}{\xi}\left(\frac{R}{h_d}\right)^3\hat{h}^{-3} - \pi\frac{\rho_v}{\rho_l}\frac{\mathcal{E}\Stn^2}{\Wen}\frac{R}{h_d} \hat{h}^{-1} \right)
  \frac{\mathrm{d}\hat{h}}{\mathrm{d}\hat{t}} - \\
  \tau^2 \frac{\gamma}{m}\left( \pi \frac{\rho_v}{\rho_l}\frac{\mathcal{E}^2\Stn^2}{\Wen}\left(\frac{R}{h_d}\right)^3\hat{h}^{-2} + \frac{6\pi}{\xi}\mathcal{E}\left(\frac{R}{h_d}\right)^5\hat{h}^{-4}\right) - \tau^2\frac{g}{h_d} = 0,
  \label{dimlessmomentum}
\end{multline}
where we introduced the damping coefficient $\beta = \eta_v R$, and the evaporation number $\mathcal{E}$ already defined in equation (\ref{evapn}).
We further simplify equation (\ref{dimlessmomentum}) by noticing that the change of momentum linked to the absolute ejection velocity, that is proportional to $1/h^2$, and the gravity term are orders of magnitude smaller than the contribution from the lubrication equation, that varies as $1/h^4$. We can therefore approximate equation (\ref{dimlessmomentum}) by:
\begin{equation}
  \frac{\mathrm{d}^2 \hat{h}}{\mathrm{d} \hat{t}^2} +
  \tau \frac{\beta}{m} \left(\frac{6\pi}{\xi}\left(\frac{R}{h_d}\right)^3\hat{h}^{-3} - \pi\frac{\rho_v}{\rho_l}\frac{\mathcal{E}\Stn^2}{\Wen}\frac{R}{h_d} \hat{h}^{-1} \right)
  \frac{\mathrm{d}\hat{h}}{\mathrm{d}\hat{t}} -
  \tau^2 \frac{\gamma}{m}\frac{6\pi}{\xi}\mathcal{E}\left(\frac{R}{h_d}\right)^5\hat{h}^{-4} = 0.
  \label{dimlessmomentum2}
\end{equation}
Equation (\ref{dimlessmomentum2}) describes a nonlinear oscillator and allows us to identify the three time scales involved in the evolution of the gas film.
We start by discussing the characteristic time linked to the $\hat{h}^{-4}$ term, that we denote $\tau_o$.
It is propoportional to $\sqrt{m/\gamma}$, indicating that it can be understood as the characteristic time of a spring-mass system.
This suggests that the lubrication force acts as a non-linear spring with time scale $\tau_o$. Plugging in typical values, we obtain $\tau_o \approx 10$ $\mu$s, a value of the same order of magnitude as the period reported in figure \ref{fig4}d.
Equation (\ref{dimlessmomentum2}) thus gives us insight into the mechanism that can lead to vertical oscillations.
If the bottom interface is perturbed towards the substrate, we compress the spring and the lubrication force increases. The interface is then repelled and it can either come back to its initial position or overshoot it. In the later situation, the lubrication pressure decreases and the interface moves down starting an oscillatory motion.
The appearance of these vibrations is controlled by the damping term, whose time scale is proportional to $m/\beta$. It is composed of a viscous damping term, with characteristic time $\tau_d = \xi m h_d^3/(6\pi\beta R^3)$, and an amplification term (\emph{i.e}, negative damping), with timescale $\tau_a = m\rho_l\Wen h_d/(\pi\beta\rho_v\mathcal{E}\Stn^2 R)$, associated to vapor ejection.
Using typical numerical values at impact, we find that the viscous damping timescale $\tau_d \approx 1$ $\mu$s is one order of magnitude smaller than the oscillation period $\tau_o$ and that the amplification characteristic time $\tau_a \approx 1$ s is larger than the inertio-capillary time $\tau_c$.
These findings are in contradiction with the observations reported in figure \ref{fig4}d where we observe fast oscillations modulated by a slower envelope that acts on a characteristic time on the order of 1 ms (figure \ref{fig4}d).
Yet, the qualitative interplay of damping and amplification reproduces some of our observations.
For a fixed thickness $h_d$, viscous dissipation always dominates at low substrate temperatures, but, as the superheat increases (\emph{i.e}, $\mathcal{E}$ increases), the damping coefficient decreases, leading to the appearance of damped oscillations and finally to unbounded oscillations as the total damping term becomes negative. This behavior is in qualitative agreement with the existence of a temperature threshold, here $T_s = 150$$^\circ$C, below which we do not observe oscillations (figure \ref{fig4}a).
When fixing the superheat $\Delta T$ and investigating the role of the film thickness, viscous damping becomes dominant as $h_d$ decreases. This evolution contradicts our observations which show that with increasing impact velocity, that is decreasing film thickness, damped oscillations appear first, followed by unbounded ones (figure \ref{fig4}a).
Our overestimation of viscous dissipation likely results from the assumption of a simplified, time-independent geometry.
Similarly as for static Leidenfrost drops, we expect the detailed interplay of viscous dissipation and evaporation to depend on the film geometry \citep{sobac2014}.

Although the minimal model does not correctly capture the detailed balance of viscous dissipation and evaporation, it provides a mechanism for oscillations that we can test against our observations.
We assume in the following that damping and amplification act on a longer timescale than the fast oscillations, as observed experimentally. We introduce this hypothesis in equation (\ref{dimlessmomentum})
by choosing $\tau_o$ as characteristic time and taking the ratio $\varepsilon = \tau_o/\tau_d$ as the fundamental small parameter of the problem:
\begin{equation}
  \frac{\mathrm{d}^2 \hat{h}}{\mathrm{d} \hat{t}^2} +
  \varepsilon\left(\hat{h}^{-3} - \frac{\tau_d}{\tau_a}\hat{h}^{-1} \right)
  \frac{\mathrm{d}\hat{h}}{\mathrm{d}\hat{t}} -
   \hat{h}^{-4} = 0.
  \label{smalleq}
\end{equation}
We separate the fast oscillations from the slow envelope using a multiscale approach \citep{hinch1991}. We define a fast time $\hat{t}_0 = \hat{t}$ and a slow time $\hat{t}_1 = \varepsilon \hat{t}$ and introduce the asymptotic expansion $\hat{h}(\hat{t},\varepsilon) = \hat{h}_0(\hat{t},\hat{t}_1) + \varepsilon\hat{h}_1(\hat{t},\hat{t}_1) + O(\varepsilon \hat{t})$ into equation (\ref{smalleq}).
At leading order, we obtain an equation that describes the dynamics of the gas film on the fast timescale:
\begin{equation}
  \frac{\partial^2 \hat{h}_0}{\partial \hat{t}^2} -
   \hat{h}_0^{-4} = 0.
  \label{lo}
\end{equation}
We now look for oscillations as a perturbation to the film thickness.
We define a perturbation of the form $\hat{h}_0(\hat{t},\hat{t}_1) = \hat{H}_0(\hat{t},\hat{t}_1) + \zeta \hat{H}_0'(\hat{t},\hat{t}_1)$ where $\zeta \ll 1$ and $\hat{H}_0(\hat{t},\hat{t}_1)$ is a solution of equation (\ref{lo}) that we assume to verify the initial conditions set at the end of the initial approach.
At leading order, equation (\ref{lo}) becomes an harmonic oscillator:
\begin{equation}
  \frac{\partial^2 \hat{H}'_0}{\partial \hat{t}^2} +
   \frac{4}{\hat{H}_0^5}\hat{H}'_0 = 0.
\end{equation}
which admits oscillatory solutions with pulsation $\hat{\omega} = \sqrt{4/\hat{H}_0^5}$.
Physically, the film thickness after the initial approach varies between the constant dimple height $h_d$ and the increasing thickness at the neck $h_n(t)$. As the oscillations appear at the dimple, we choose $h_d$ as a typical scale for the initial film thickness. With this choice, we obtain the oscillation frequency:
\begin{equation}
  f_{th} = \frac{1}{2\pi}\sqrt{\frac{24\pi}{\xi}\frac{\gamma}{m}\mathcal{E}\left(\frac{R}{h_d}\right)^5} = \frac{1}{2\pi\tau_c}\sqrt{ \frac{18}{\xi}\mathcal{E}\left( \frac{R}{h_d} \right)^5 }.
  \label{freq}
\end{equation}

\begin{figure}
\centering
  \includegraphics[width=1\textwidth]{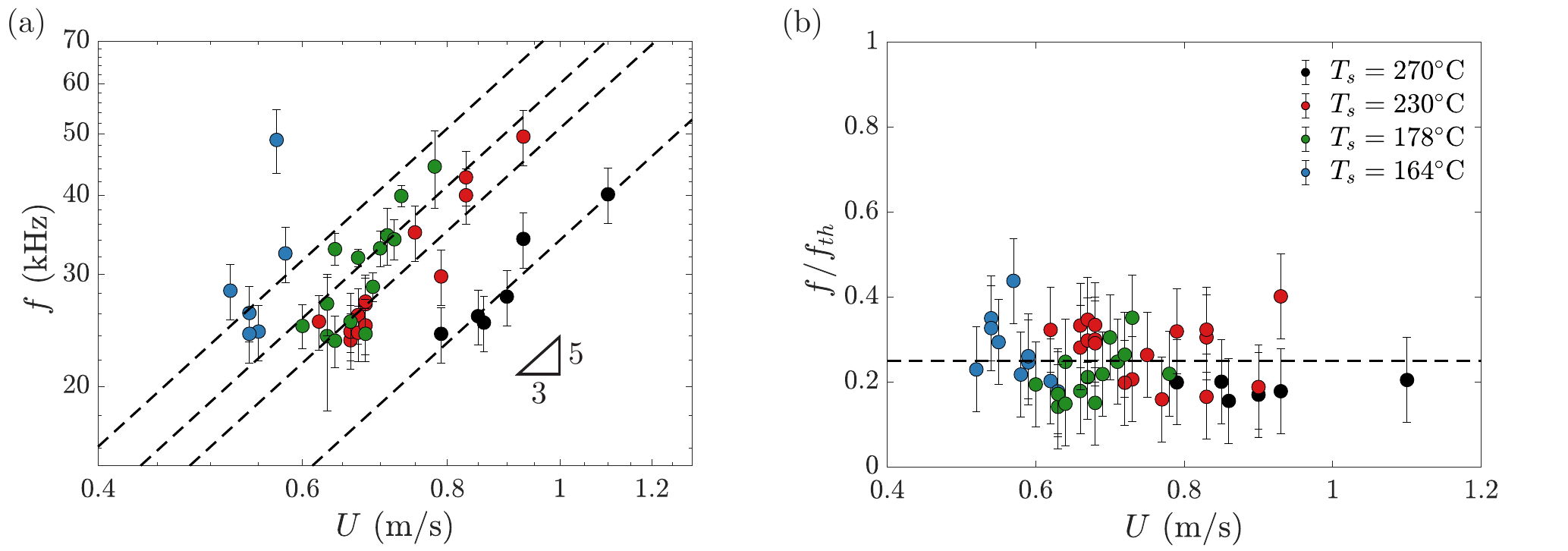}
  \caption{(a) Oscillation frequency $f$ of the gas film for impacts at velocity $U$ on substrates with different $T_s$ (same legend as in (b)). Dashed lines with slope $5/3$ are drawn as a guide to the eye. (b) Measured frequency $f$ divided by the frequency $f_{th}$ predicted from equation (\ref{freq}) as a function of $U$ for various substrate temperatures. The data collapse on a constant value $0.25 \pm 0.1$. Although the data is consistent with the scaling of equation (\ref{freq}), the prefactor in that expression is about four times too large.}
\label{fig5}
\end{figure}

We now compare the observed value of $f$ to the prediction of equation (\ref{freq}).
In figure \ref{fig5}a, we show the measured frequency $f$, extracted from the space-time representation of the RI data (figure \ref{fig4}d), as a function of the impact velocity $U$ for various substrate temperatures $T_s$.
The frequency $f$ ranges from $22$ kHz to $48$ kHz.
For a fixed superheat, $f$ increases with the impact velocity. This variation is in qualitative agreement with the prediction of equation (\ref{freq}) as, combining it with the scaling relation linking the dimple height to the Stokes number (equation (\ref{hd})), we obtain a power-law relationship with exponent $5/3$ between $f_{th}$ and the impact velocity $U$, or in dimensionless form, the Stokes number \Stn:
\begin{equation}
  f_{th}\tau_c \sim \Stn^{5/3}.
\end{equation}
This power-law behavior is consistent with the experimental variation displayed in figure \ref{fig5}a, where the dashed lines are drawn as guides to the eye with a slope $5/3$.
The effect of the substrate temperature $T_s$ is more subtle as it affects both the evaporation number $\mathcal{E}$ and the dimple height $h_d$.
To quantitatively test the frequency prediction, we need to evaluate the temperature dependant gas properties. As the conduction timescale $h_d^2/\kappa_v$ is on the order of $0.1$ $\mu$s, we consider steady heat transfer in the vapor film.
We evaluate the density, thermal conductivity and viscosity at $(T_s+T_b)/2$ and plot the measured frequency $f$ normalized by the prediction $f_{th}$ as a function of $U$ (figure \ref{fig5}b) where we measure $h_d$ and we use $\xi = 4$ as the ratio of the liquid to the vapor viscosity is unusually low, $\eta_l(T_b)/\eta_v((T_s+T_b)/2) \approx 40$, allowing slip at the liquid-gas interface.
The data collapse on a constant $0.25 \pm 0.1$ when varying $T_s$. The hydrodynamic model captures both the effect of superheat and impact velocity, but consistently overpredicts by a factor four the oscillation frequency.
This discrepancy can be explained by the simplicity of the model that disregards the spatial variation of the gas film thickness and the temporal variation of its radial extent.

\section{Conclusion and outlook}
We have studied the impact of volatile drops on superheated substrates, revealing new features of the gas film dynamics and of the Leidenfrost transition.
First, using high-speed interferometry, we disentangled the role of gas drainage and evaporation on the short-time dynamics of the gas layer. We found that the superheat noticeably affects the vertical position of the liquid-gas interface.
The role of vapor production is dominant at the neck, close to the hot solid, where the gas pressure is set by a balance of inertia and evaporation. Ultimately, this pressure is balanced by the capillary pressure as the interfacial curvature increases, setting the minimum film thickness.
On the contrary, the initial drop deformation, that is dimple formation, is determined by a balance of gas drainage and liquid inertia, similarly as for isothermal impacts.
The subtle role of evaporation suggests limits to our description. Liquids with markedly different thermal properties could have different balances between drainage, evaporation and further effects. Particularly, liquids with low latent heat could extend the influence of vapor generation to dimple formation. Also, performing experiments in vacuum would provide insight into the limit where only evaporation contributes to levitation.

Second, we showed that the dynamic Leidenfrost transition is affected by both gas drainage and evaporation. (i) The transient stability of the draining gas film can enable drop rebound, remarkably resulting in a dynamic Leidenfrost temperature lower than its static value in the low impact velocity limit. (ii) We found a hitherto unreported collapse mode of the gas layer specific to impact on superheated substrates. For large superheat, contact at the transition to the dynamic Leidenfrost effect is induced by vertical axisymmetric oscillations of the drop's bottom interface that are the result of the coupling of drop motion and of the gas flow generated by evaporation.
The transition to contact cannot be understood by only accounting for the minimum film thickness: multiple mechanisms can trigger wetting during the different timescales associated with drop impact.
The minimal hydrodynamic model we propose, while accounting for the influence of impact velocity and substrate temperature, does not allow to predict the transition to contact. We expect that predicting the threshold for the appearance of oscillations would require to take into account the shape of the gas film, a task that is beyond the scope of the present study.

\section*{Acknowledgments}
We thank Ambre Bouillant, Pallav Kant, and Jos\'e Manuel Gordillo for fruitful discussions and acknowledge funding from the ERC Advanced Grant DDD under grant-\# 740479. 
\appendix
\section{TIR and RI image processing and calibration}
\label{appendix_setup}
\subsection{Image analysis}
\subsubsection{TIR}
TIR image processing has been described in details by \citet{shirota2017}, we recall here the main steps of the analysis.
(i) First, we compensate for the optical transformation of the setup: a circular object becomes an ellipse with a principal axis ratio $D_S/D_L$ that is related to the angle of incidence $\phi$ of the laser beam. We report here the modified function linking $D_S/D_L$ to $\phi$, a consequence of the presence of the optically coupled sapphire window:
\begin{multline}
  \frac{D_S}{D_L} = \frac{\cos\left( \sin^{-1}\left( \frac{n_s}{n_{g}}\sin\phi \right)\right)}
  {\cos\left(\sin^{-1}\left(\frac{n_s}{n_{g}}\sin\phi\right)  -\frac{\pi}{4}\right)}
  \cos\left(\sin^{-1}\left(n_{g}\sin\left(\sin^{-1}\left(\frac{n_s}{n_{g}}\sin\phi\right)-\frac{\pi}{4}\right)\right)\right)
\end{multline}
where $n_s$ and $n_g$ are the optical indices of the sapphire window and of the glass prism respectively.
(ii) The transformed images are then divided by the background image to obtain intensity normalized snapshots.
(iii) Finally the grayscale images can be converted into absolute height fields using a lookup table relating the normalized intensity to the height.
\begin{figure}
  \centering
  \includegraphics[width=\textwidth]{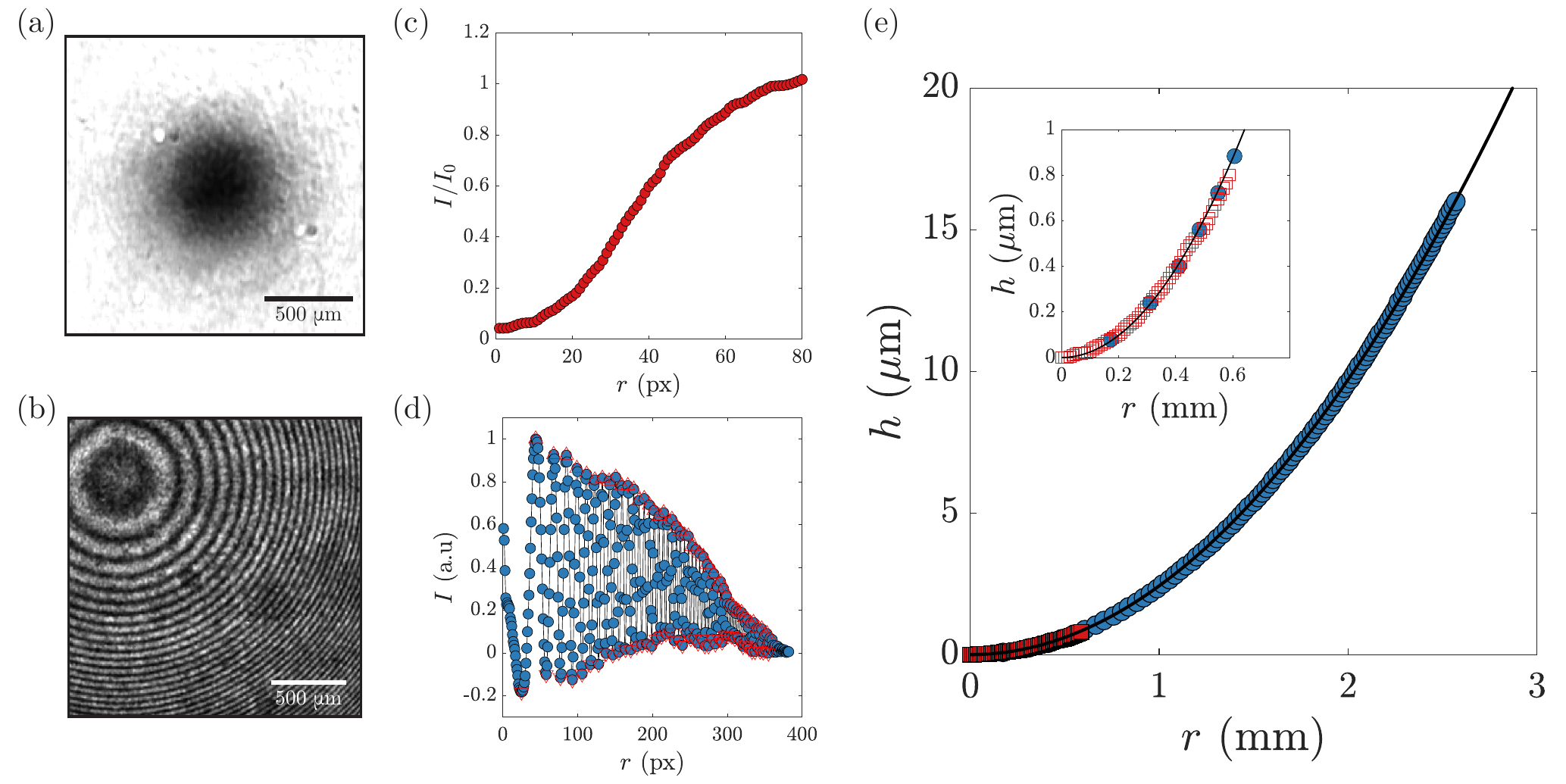}
  \caption{\label{calib}(a) TIR image of the gap between the lens and the substrate after transformation and intensity normalization. (b) Interference pattern generated by the gap between the lens and sapphire disk (c-d) Radial intensity profiles obtained from a and b. The red triangles highlight the detected peaks in d. (e) Height of the interface as a function of the radial coordinate $r$ as determined from TIR (red squares) and RI (blue dots) and compared to the lens profile (black solid line). The inset zooms on the evanescent lengthscale where TIR and RI measurements overlap.}
\end{figure}
\subsubsection{RI}
We process RI images by performing background substraction and azimuthal averaging, taking advantage of the axisymmetry of the problem, to improve the signal to noise ratio.
The interference patterns obtained at each instant are then transformed into 1D intensity profiles that can be stacked in space-time diagrams that allow to visualize the impact dynamics. Each fringe is an iso-height line and adjacent bright (dark) fringes corresponds to a thickness variation $\lambda/2$. We use standard peak detection algorithms to detect the peaks in the 1D profiles and track their position in time using particle tracking. This representation enables to visualize height contours in space and time, a property that allows one to determine the absolute thickness at any time from its knowledge at a single instant.

\subsection{Calibration}
We calibrate the setup before each set of experiments by measuring the gap between a concave lens with known radius of curvature ($R_{lens} = 206$ mm) and the sapphire disk.
Figures \ref{calib}a and b respectively show the TIR and RI images after data processing. From these, we extract azimuthally averaged intensity profiles (Figures \ref{calib}c and d).
In figure \ref{calib}e, we plot the gap thickness between the lens and substrate as a function of $r$ as determined from TIR (red squares) and RI (blue dots) and compare it to the known lens profile (black solid line). The inset enables us to visualize the overlap between TIR and RI measurements that allows us to obtain the absolute film thickness from $\sim 10$ nm to $\sim 10$ $\mu$m by matching the absolute information from TIR with the relative information of RI.

We estimate the error in TIR measurements from the calibration curve. The error is mainly caused by the uncertainty in the determination of the incidence angle and the steepness of the function relating large heights to intensity. The error is on the order of 10 nm for small heights and increases as we get close to the upper bound of the measurement range reaching typically 100 nm.
For RI measurements, the main source of uncertainty comes from the limited resolution that could prevent to distinguish individual peaks in the dimple region. The error bars are an upper bound that corresponds to the blending of two successive black (white) fringes.

\section{Minimum thickness for isothermal impacts}
\label{appendix_hmin}
We derive the scaling for the minimum thickness of the gas film in isothermal conditions, obtained by \citet{mani2010}. We start from the self-similar set of governing equations, (\ref{selfsimliquid}) and (\ref{selfsimgas}), where the right hand side term of (\ref{selfsimgas}) is set to zero.
In isothermal conditions, the viscous lubrication flow opposes the liquid inertia at the neck. Combining (\ref{selfsimliquid}) and (\ref{selfsimgas}) gives the following scaling relations for the pressure $\tilde{p}_n$ and the lengthscale $\tilde{\ell}$:
\begin{equation}
  \tilde{\ell} \sim \tilde{c}^{1/2}\tilde{h}_n^{3/2}, \qquad \tilde{p}_n \sim \tilde{c}^{3/2}\tilde{h}_n^{-1/2}.
\end{equation}
As $h_n$ decreases, the initially neglected Laplace pressure at the neck $\gamma h_n/\ell^2$, which scales as $h_n^{-2}$, diverges faster than the gas pressure at the neck $p_n$ which scales as $h_n^{-1/2}$, indicating that the initial hypothesis to neglect surface tension is no longer valid as the drop approaches the solid.
The balance between $p_n$ and the Laplace pressure thus sets the minimum film thickness $h_m$:
\begin{equation}
  \frac{h_m}{R} \sim \Wen^{-2/3}\Stn^{8/9}.
  \label{hmscalingiso}
\end{equation}
\begin{figure}
  \centering
  \includegraphics[width=\textwidth]{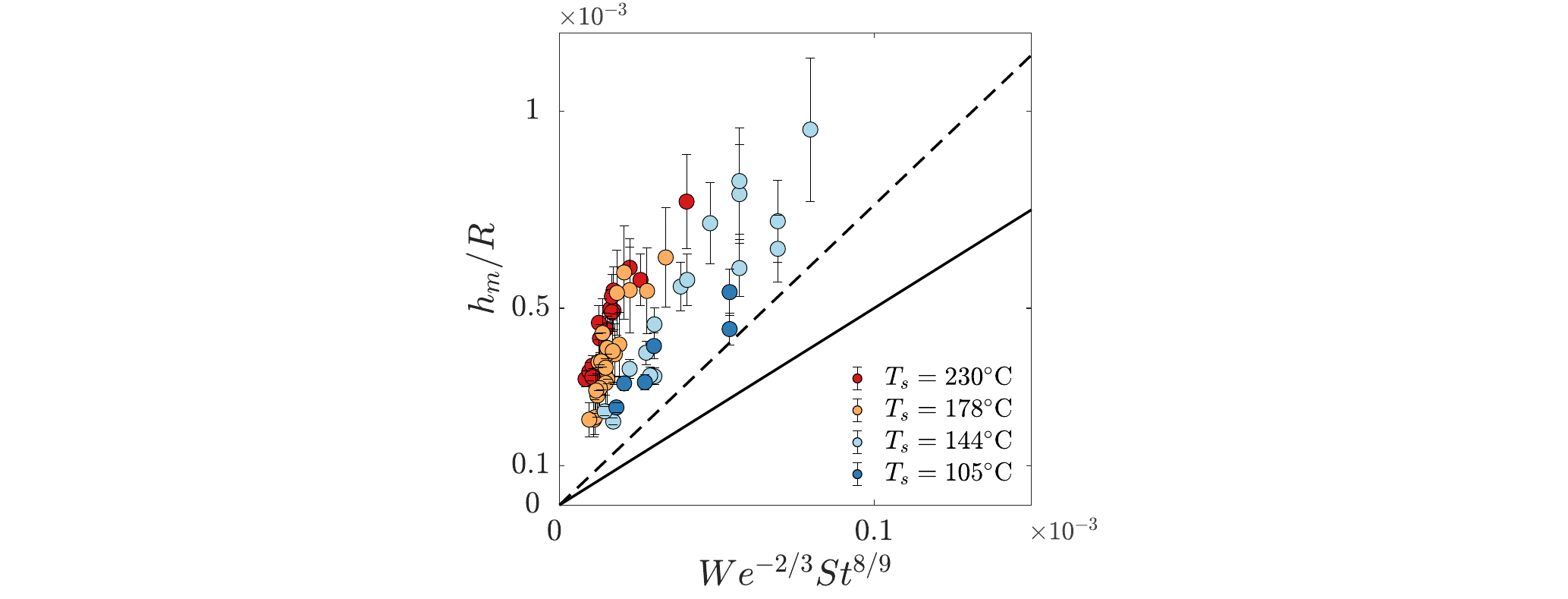}
  \caption{\label{figC} Normalized minimum film thickness $h_m/R$ as a function of the prediction of equation (\ref{hmscalingiso}). The solid line stands for the prefactor calculated by \citet{mandre2012} and the dashed line represents the measured prefactor from the measurements of \citet{ruiter2012}.}
\end{figure}
In figure \ref{figC}, we plot the measured minimum film thickness as a function of the scaling (\ref{hmscalingiso}) where the surface tension is evaluated at $T_b$ and the gas viscosity is extracted from our measurements of the dimple height $h_d$. Even if the isothermal and superheated scalings predict similar power-law decrease of $h_m$ with $U$ with exponents $-20/9$ and $-2$, respectively, the data for different substrate temperatures do not collapse on a single line. The isothermal scaling does not capture the effect of substrate heating: the measured thicknesses are always larger than that predicted numerically (\citet{mandre2012}, solid line) or reported in experiments in the isothermal case (\citet{ruiter2012}, dashed line), and the deviation increases with increasing superheat. This comparison confirms that for impacts on superheated surfaces, the liquid inertia at the neck is balanced by vapor generation.

\section{Short-time rupture of the gas film}
\label{appendix_rupture}
\begin{figure}
  \centering
  \includegraphics[width=\textwidth]{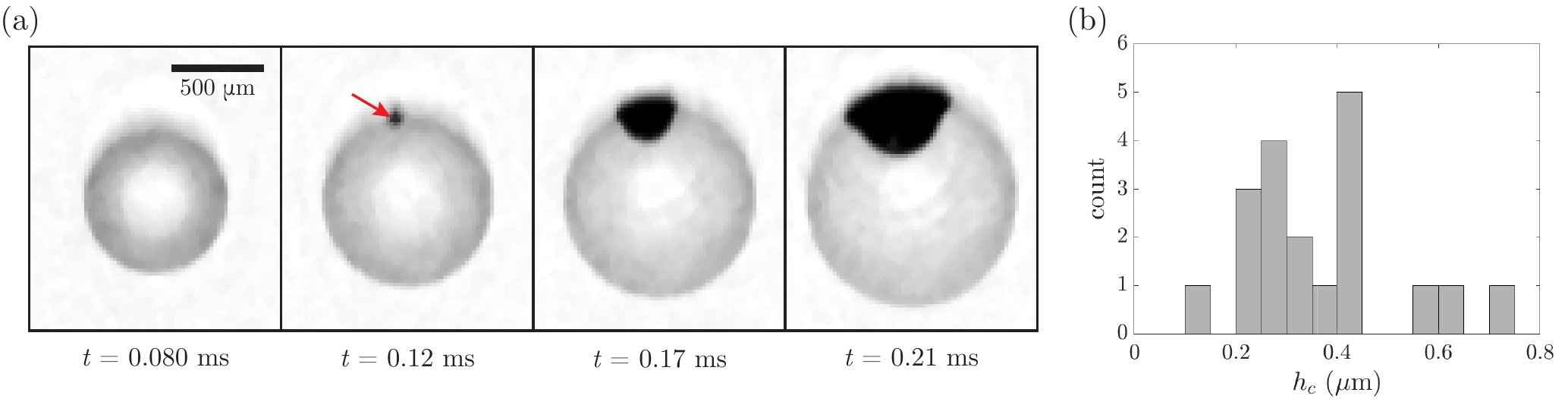}
  \caption{\label{figB}(a) TIR image sequence of contact initiation at short-time ($U = 0.5$ m/s, $\Wen = 9$, and $T_s = 144$$^\circ$C). Contact occurs at the neck at $t = 0.12$ ms (red arrow), the rupture thickness at the contact location is $h_c = 0.64$ $\mu$m.  (b) Histogram showing the value of $h_c$ in 20 experiments with $T_s$ ranging from 105 to 164$^\circ$C.}
\end{figure}

We monitor the nucleation of point-like wetting spots that precede the collapse of the gas film.
Figure \ref{figB}a shows a sequence of TIR grayscale images associated to an impact with $U =0.5$ m/s (\emph{i.e} $\Wen = 9$) and $T_s = 144$$^\circ$C. The drop spreads on a gas cushion before a contact spot nucleates at $t = 0.12$ ms (red arrow).
We measure the time evolution of the thickness at the location where contact occurs to obtain the film height before rupture $h_c = 0.64$ $\mu$m, that is larger than the global minimum of the thickness at the same time instant, here $0.44$ $\mu$m.
We are able to resolve $h_c$ only for moderate impact velocities as sufficient time resolution is needed to distinguish the rupture from the thinning dynamics.
In figure \ref{figB}b, we plot the histogram of $h_c$. The film rupture height is broadly distributed, with an average value of $0.36$ $\mu$m, in agreement with data obtained for isothermal impacts on glass \citep{ruiter2012,kolinski2014_2}, suggesting that the presence of defects on the surface triggers liquid-solid contact.
During impact on heated substrates, contact always occurs at the neck at short-time unlike what is observed for ambient temperature surfaces. The film shape promotes this location: there is no thin  extended region bridging the dimple to the outer edge of the film.

After nucleation of contact, the wetting spot spreads ($t = 0.17$ ms and $t = 0.21$ ms), preferentially in the azimuthal direction along the neck, binding the drop to the substrate similarly as during isothermal impacts. These wetting dynamics disappear when we increase $T_s$ as we observe the formation of boiling patterns immediately after contact (Figure \ref{fig4}b).

\bibliographystyle{jfm}
\bibliography{Bibli}
\end{document}